\documentclass[opre]{informs3nothing}
\pdfoutput=1

\usepackage{amsmath}
\usepackage{amssymb}
\usepackage[numbers]{natbib}
\NatBibNumeric

\renewcommand{\v}[1]{\mathbf{#1}}
\newcommand{\vx}{\mathbf{x}}
\renewcommand{\P}{\mathbb{P}}
\newcommand{\E}{\mathbb{E}}

\newcommand{\vu}{\mathbf{u}}
\newcounter{thm}
\newtheorem{mythm}[thm]{Theorem}
\newcommand{\qcount}[1]{
	\left\lceil\frac{N_{#1}}{N_{M}}\right\rceil
}

\TITLE{Finding Online Extremists in Social Networks}
\RUNTITLE{Finding Online Extremists in Social Networks}

\date{\today}

\RUNAUTHOR{Klausen, Marks, and Zaman}
\ARTICLEAUTHORS{%
\AUTHOR{Jytte Klausen}
\AFF{%
Brandeis University \\ 
415 South Street \\
Waltham, MA~ 02453 \\ 
\EMAIL{klausen@brandeis.edu}
}
\AUTHOR{Christopher E Marks}
\AFF{%
Operations Research Center, Massachusetts Institute of Technology\\
Charles Stark Draper Laboratory \\
555 Technology Square \\
Cambridge, MA  02139 \\
\EMAIL{cemarks@mit.edu}
}
\AUTHOR{Tauhid Zaman}
\AFF{%
Sloan School of Management, Massachusetts
Institute of Technology \\ 
77 Massachusetts Ave. \\
Cambridge, MA 02139 \\ 
\EMAIL{zlisto@mit.edu}
}

} 

\ABSTRACT{
Online extremists in social networks pose a new form of threat to the general public.  These extremists range from cyberbullies who harass innocent users to terrorist organizations such as the Islamic State of Iraq and Syria (ISIS) that use social networks to recruit and incite violence.  Currently social networks suspend the accounts of such extremists in response to user complaints. The challenge is that these extremist users simply create new accounts and continue their activities.  In this work we present a new set of operational capabilities to deal with the threat posed by online extremists in social networks.
 
Using data from several hundred thousand extremist accounts on Twitter, we develop a behavioral model for these users, in particular what their accounts look like and who they connect with.  This model is used to identify new extremist accounts by predicting if they will be suspended for extremist activity.  We also use this model to track existing extremist users as they create new accounts by identifying if two accounts belong to the same user.  Finally, we present a model for searching the social network to efficiently find suspended users' new accounts based on a variant of the classic Polya's urn setup.  We find a simple characterization of the optimal search policy for this model under fairly general conditions. Our urn model and main theoretical results generalize easily to search problems in other fields. }


\begin{document}

\maketitle


\section{Introduction} \label{intro}


In recent years there has been a huge increase in the number and size of online extremist groups using social networks to harass users, 
recruit new members, and incite  violence.  
These groups include terrorist organizations such as the Islamic State of Iraq and Syria (ISIS) \cite{berger2015}, white nationalists and Nazi sympathizers \cite{ref:nazi}, and cyberbullies who target individuals with offensive and harassing
messages \cite{ref:milo}.  Of particular concern is the danger posed to public safety by terrorist groups.    The  threat from terrorist groups such as ISIS has become so severe that U.S. president Barack Obama recently said ``The United States will continue to do our part, by working with partners to counter ISIL's\footnote{ISIL is another name for ISIS and stands for is the Islamic State of Iraq and the Levant.} hateful propaganda, especially online" \cite{ref:obama_isis}.  It is suspected that the online presence of ISIS may  have been
responsible for radicalizing individuals and motivating them to commit acts of terror \cite{ref:orlando}.  

Social network  have recently begun taking actions to actively combat online extremists.
  For instance, Twitter, which has become
the main venue for ISIS users to spread their propaganda \cite{ref:obama_isis}, 
has been very aggressive in its response to ISIS.
In August 2016, Twitter reported that it had shut down over 360,000 ISIS accounts and  its daily suspensions of terrorism-linked accounts have jumped 80 percent since 2015 \cite{ref:twitter_isis_shutdown}.  Twitter identifies extremist
accounts primarily based on reports from its users, but it has begun using    
proprietary spam-fighting tools to supplement these reports.  
These tools have helped to automatically identify more than one third of the accounts that were
 ultimately suspended for promoting terrorism on Twitter \cite{ref:twitter_spam_isis}.  

The efforts of social networks such as Twitter have been effective at limiting the reach of
online extremist groups such as ISIS.  However, not all extremist users are shut down
and they are constantly returning to the social network after being suspended.  In addition,
much of the success in mitigating the threats of extremist groups has relied upon
the cooperation of the social networks themselves.  For instance, Twitter has dedicated teams to review
user reports of potential extremist accounts \cite{ref:twitter_isis_shutdown}.
However, if extremist users migrate to other social networks, there is no guarantee
that the companies which operate these networks will be as cooperative or dedicate
as many resources to dealing with online extremists.  Therefore, what is needed is a set of capabilities
that can be used by authorities to combat online extremists which do not rely upon the cooperation
of the social network operators and can be applied to any social network.

\subsection{Our Contributions}
The case of ISIS in Twitter is useful to understand general behavioral
patterns of online extremist users in social networks.  We  use these behaviors to
guide the development of capabilities for combating online extremists in general social networks.  We provide
a detailed analysis of these behaviors and develop the corresponding capabilities
 in Sections \ref{suspensions}, \ref{similarity}, \ref{refollowing}, and \ref{search}.
Here we will provide a concise overview of our major contributions, in particular the
different behavioral patterns of online extremists and the corresponding capabilities we develop.

\textbf{Suspensions.}  Online extremist users post content which violate the Terms of Use of social networks, leading to
the suspension of their accounts.  These suspensions occur in response to user reports, but 
many social networks are beginning to use  algorithms to automatically detect any violative content.
Going one step further, it would be useful to have a capability to flag users as potential extremists
before they post any content at all.  There are potential features of an account that may predict if
it belongs to an extremist user.  For instance, the account may not publicly declare its geographical location.
Also, the users to which the account connects may indicate whether or not the account belongs to an extremist user.
In Section \ref{suspensions} we use these intuitions to develop a method to automatically predict if an account will be suspended
without requiring it to post any content.

\textbf{Creating Multiple Accounts.}  After being suspended, online extremist users will quickly
create new accounts and continue their activities on the social network.  This makes it difficult
to keep an extremist user off the social network.  Typically the new account resembles the suspended
account in several aspects.  For instance, the names and profile pictures may be very similar.
A useful capability would be the ability to identify if multiple accounts as belong to the same user.
This would allow for more accurate monitoring and tracking of extremist users.  
We develop such a capability in Section \ref{similarity}.

\textbf{Refollowing Previous Friends.}  A user in a social network generally follows
a set of users. In Twitter these followed users are referred to as the \emph{friends} of the user and the user
is referred to as their \emph{follower}.  Upon returning to the social network after being suspended,
an extremist user will generally refollow  some of his previous friends.
  If we knew which previous friends a suspended user refollows, 
	this information could be used to find the user's new account in the social network.
There may be features of the friends which make it more likely the suspended user
will refollow them.  In Section \ref{refollowing} we use these features to develop
a method to predict who suspended users refollow.

\textbf{Suspended User Search.}  
Authorities may wish to find suspended users when they return 
to a social network.  The operator of the social network
is notified every time a new user enters the network
and can use our account matching capability to see if the
new user matches a previously suspended user.  However, if one
is not the operator of the social network, then one must search
the network to see if the suspended user has returned.  Because
of the size of the social network, this search
could require a large amount of time and resources.  To overcome
this challenge, we develop
an efficient network search policy in Section \ref{search} based
on a variant of a Polya's urn model
which utilizes our refollowing prediction capability
from Section \ref{refollowing}.

The remainder of this paper is organized as follows.
We review the extant literature relevant to our work in Section \ref{sec:previous_work}.
We provide a detailed overview of the data used for our analysis in Section \ref{data}.
Section \ref{suspensions} presents our predicting suspensions
capability.  We present our account matching capability in Section
\ref{similarity}.  Our method for predicting refollowing is presented
in Section \ref{refollowing}.  Section \ref{search} details our model for network
search and an optimal search policy.  We conclude in Section \ref{conclusion}.

\subsection{Previous Work}\label{sec:previous_work}

\textbf{Analysis of Online Extremist Networks}
There are several studies focused on ISIS users in social networks.
One of the first studies characterizes the number,
behavioral traits, and organization  of Twitter ISIS users \citep{berger2015}.
A subsequent study by the same authors found that the reach 
of ISIS had been limited by the beginning of 2016 due to
the efforts of Twitter to suspend ISIS accounts \cite{berger2016islamic}.
In  \cite{johnson2016new} the authors study the dynamics of ISIS users in the Russian social
network VKontakte and suggest that shutting down smaller pro-ISIS groups
can prevent the emergence of larger, more influential groups. In \cite{ferrara2016predicting}
the authors develop models to predict which users will be suspended for being in ISIS,
who will retweet ISIS content, and who will interact with ISIS users.  This work is similar to our
work, but the authors do not study many of the capabilities we develops such as
identifying multiple accounts from a single user, refollowing
old friends, or searching for suspended users.

There have also been several works looking at identifying extremist
content in groups beyond ISIS.  In  \citep{scanlon2014automatic} the authors develop
methods to automatically classify content that is used for recruiting
members to extremist groups.  Similar work in \cite{sureka2014learning}
used machine learning to detect content that promotes hate and extremism.
Machine learning methods have also been used to detect
cyberbullies based on the content they post \cite{reynolds2011using,dinakar2011modeling}.
The work in \cite{dinakar2012common} builds upon this work to develop
an approach for mitigating the threat of cyberbullying.

\textbf{Spam/Bot Detection}
Closely related to our capabilities on predicting suspensions is the work
done on detecting online bots (non-human users) or malicious users.  
Several approaches have been developed which use different types of behavioral features.
The type of content
(URL's, user mentions) was found to be predictive of Twitter bots in \cite{lee2011seven}.
  Temporal behavior and aggregate network properties (in-degree,out-degree) were used to identify Twitter bots in \cite{chu2012detecting}.  
In \cite{dickerson2014using} the authors demonstrate that the sentiment of the posted
content can be used to identify bots.   All of these approaches are designed to detect automated behavior.  However, they
may not be as effective for human users who engage in extremist behavior.  Also, many of these approaches
require the user to post some content in order to detect whether or not they are bots.
An approach related to ours is in \cite{kumar2014accurately} which relies
purely on network structure to identify malicious users in social networks where edges have a polarity (friend/enemy).
In contrast to this extant work, our approach combines both behavioral features with refined network features
to detect extremist users.

\textbf{Network Search}
Our  network search problem is similar to those presented by \citet{alpern2013mining} and \citet{NET:NET20241}, who have done much work in this area.  Unlike their work, in which the searcher and the target are assumed to be operating in a physical network, our problem of searching a social network admits a different set of search constraints.  In our network search problem, the searcher is not constrained to move along edges.  Instead, the searcher can examine the neighbors of any of a set of nodes that are known to him, but each of these queries comes at a cost.
This alternative representation of network search follows from one of the original search problems posed by \citet{black1965discrete}, in which a searcher looks for the search target among a set of possible locations.  Each location has a known probability of containing the target and a known probability of finding the target, if it is there.  Our network search application adapts this simple search model to a network setting.  Instead of limiting the search target to be at at most one of a set of possible locations, in a network search the target could be connected to more than one of the nodes known to the searcher.  Also, the method of querying the neighbors of a node causes the probability of finding the target to change with each observation. 

Our network search model builds directly on the multi-urn search model presented in \cite{urn1}.  However, in this work the major difference is that we allow for more than one query
to be done in each step, which results in slight differences in the optimal policies.

\section{Data}\label{data}
The data we study in this work comes form the micro-blogging site
Twitter \cite{twitter}. 
Twitter serves as a front line public platform used by ISIS for outreach and recruitment.  ISIS's presence on Twitter, and its consistent success at gaining support and recruits through the social media site has been deeply analyzed and well-documented \cite{berger2015}.  

Twitter users form a social network by connecting to each other.  This network is directed and this directionality dictates the flow of information.  A user forms a connection with someone on Twitter by \emph{following} him or her.  Each account a user is following is known as this user's \emph{friend} and the user is known as the friend's \emph{follower}.  These friends/followers edges form the Twitter social network.  This network is then used to transmit information.  In Twitter, this information comes in the form of short messages that  users post known as \emph{tweets}.  When a user posts a tweet, that tweet appears in the Twitter timeline of all the user's followers.  In this manner, information flows from users to their followers in Twitter.

For this research, we collected Twitter data from approximately 5,000 ``seed'' users, who were either known ISIS members  or who were connected to many known ISIS members as friends or followers.  The names of these seed users were obtained through news stories, blogs, and reports released by law enforcement agencies and think tanks \cite{klausen2015tweeting}.  The data was collected at various times throughout the calendar year 2015, using Twitter's REST API (see \cite{rest}).

For each seed user we collected the user account profile information, including the screen name, name, description, location, profile picture, and profile banner at the time of the collection.  We also obtained the user account ID number, which is the only unchangeable unique account identifier. In addition to obtaining seed users' profile information, we collected the same set of profile information for each  seed user's friends and followers.  As a result the number of user profiles contained in the data set grew to over 1.3 million.

We downloaded all publicly available tweets from each seed user's timeline at the time of collection.  For each tweet we obtained the unique tweet ID assigned by Twitter, the tweet text, the time of the post, all hashtags, user mentions, URLs, and images contained in the tweet, and whether the tweet was a retweet of or reply to another tweet.  The total number of tweets in our data is approximately 4.8 million.

Finally, we tracked many of the accounts for several months in 2015 in order to see if they were ever suspended.
We do not know the reason for suspension, but given that these accounts were associated with known ISIS users, we assume
the suspension was related to some form of extremist propaganda that violated Twitter's user agreement.  We
tracked all of the user accounts collected in June, 2015, including the seed accounts and their friends' and followers' accounts. This data set includes 646,961 accounts in total, of which 35,080 (or 5.4\%) had been suspended as of September 23, 2015.


\section{Predicting Account Suspensions} \label{suspensions}
The first capability we develop to combat online extremism is to predict which accounts
belong to new extremist users.    In this section we develop an approach to this using
logistic regression.  We label any account in our data set as extremist if it was suspended by Twitter.
Therefore, to detect extremists we predict which accounts are suspended by Twitter.  We accomplish this
using a logistic regression model based upon features of the user accounts.
 We provide out-of-sample performance  evaluation of the model and provide insights on what factors might be useful in predicting whether a Twitter user is going to be suspended for violative behavior.


To train, validate, and test this prediction model we use a subset of the accounts whose suspension status
we tracked.  We randomly selected two non-overlapping samples of this data sets, each consisting of 5,000 accounts and maintaining the 5.4\% suspension rate, which is the overall suspension rate of these accounts.  These data sets were used for training and validation.

For our logistic regression model, the response variable is whether the account was still active as of September 23, 2015.
The predictors are obtained from the  wide array of information associated
with the user accounts.   Some of these relate to the account itself, while others
have to do with the network connections of the account.  The variables used as predictors for our model
are listed in Table \ref{table:FeaturesSuspensions}.  While we have observed that the number of screen name changes associated with a user account might serve as a good predictor of future suspension, we assume that this information is not necessarily known for an arbitrary account we wish to classify.  Similarly, we assume we do not know if the account was following accounts that were suspended in the past.  All features we use are what could be measured for a new account that
has not been seen before.  
\begin{table}
	\centering
		\begin{tabular}{|c|l|}
		\hline
		Feature type & Feature \\ \hline
		Network & Following each of 2,376 active ISIS seed accounts in our data \\
		        &(2,376 binary variables). \\ \hline
		Account & Date and time the account was created (numeric).\\ \hline
Account &Number of ``friends'' and ``followers'' connected to the account \\ 
        &(2 numeric variables).\\ \hline
 Account &Number of tweets  from the account (numeric).\\ \hline
 Account &Geo-location enabled (binary).\\ \hline
 Account &``Protected'' account (posts are not visible to the public) (binary).\\ \hline
 Account &Verified account (identity confirmed by Twitter) (binary). \\ \hline
		\end{tabular}
	\caption{Features for predicting Twitter suspensions.}
	\label{table:FeaturesSuspensions}
\end{table}

\subsection{Results}

We fit a logistic regression model with $L_{1}$-norm regularization to the training data.  From validation, we find that setting the regularization constant to 10 consistently provided near-optimal performance.  The resulting coefficient estimates were nonzero for 89 of the predictor variables, of which 81 corresponded to following certain accounts.  The signs and magnitudes of the coefficients give us some idea of the effects of some of the predictor variables.  The coefficient estimates indicate that accounts that had enabled geo-tagging and accounts that had Twitter-verified owners were much less likely to be suspended.  This is not surprising given that we expect
online extremists to want to mask their identity and location.  
The effects of friendships were less intuitive and difficult to interpret.  In total we found 
that 38 accounts had a positive sign and 43 had a negative sign.  However, there was no
clear pattern that we could find among the positive sign accounts or negative sign accounts.
More detailed analysis may reveal what made following these accounts increase or
decrease the likelihood of suspension.  Nonetheless, just knowing the value
of the regression coefficient was sufficient to predict suspensions. 

\begin{figure}  
\centering
\includegraphics[scale=0.33]{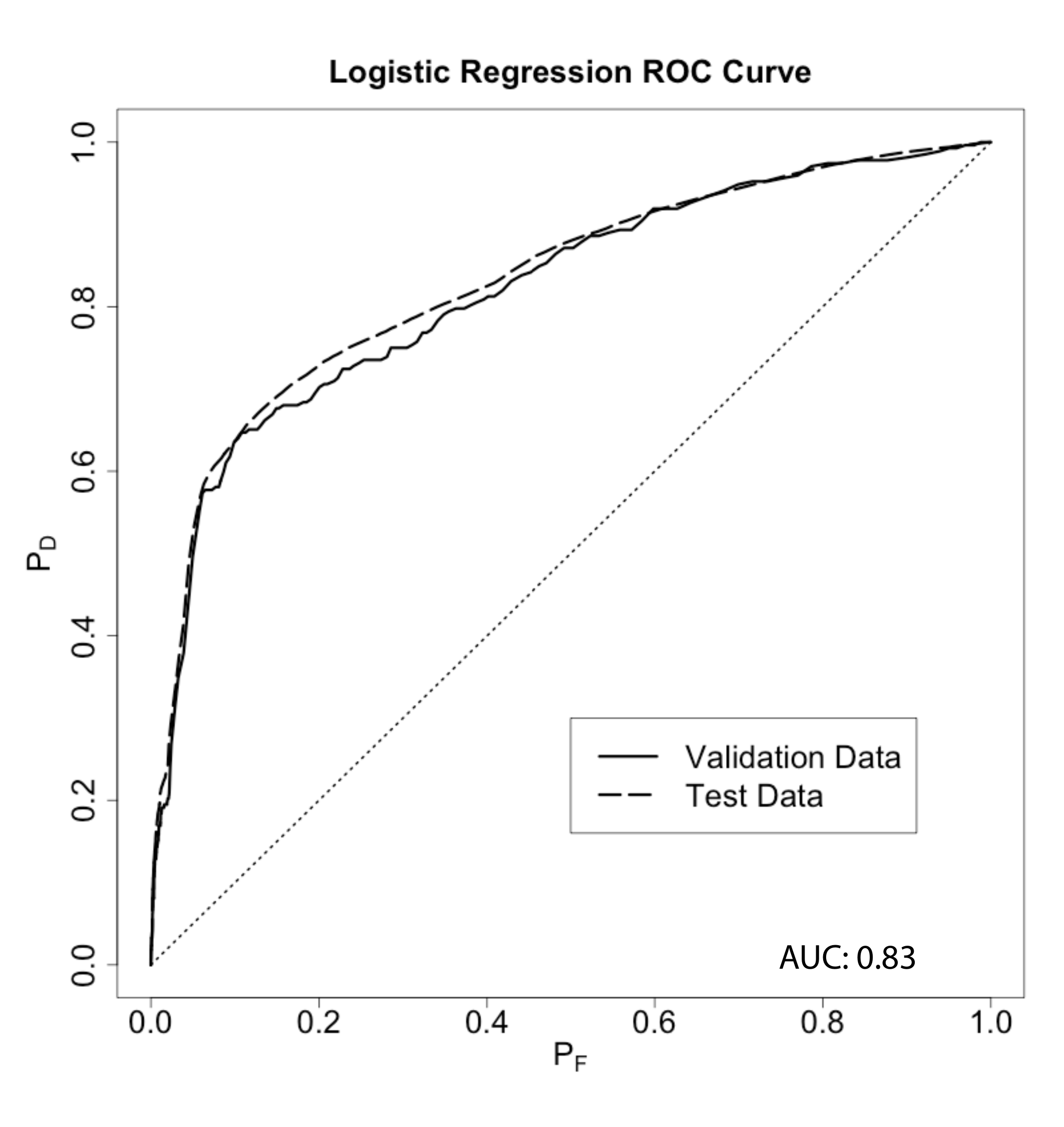}
\caption{ROC curve for the regularized logistic regression classifier for Twitter suspensions.} \label{roc} 
\end{figure}

Figure \ref{roc} shows the receiver-operator characteristic (ROC) curve on the validation set and on the test data, which was comprised of the 636,961 accounts not used for training and validation.  The area under the curve (AUC) on the test data is approximately 0.83.  We can see from the curve that we can detect about 60\% of suspended users in the test set with only a 10\% ``false positive'' rate.  This efficiency occurs by setting a classification probability of 0.1, i.e., classifying an account as an ISIS account if the regression function assigns a probability of suspension higher than this threshold.
It is important to note that because 94.6\% of the accounts in our data were not suspended, a 10\% false positive rate represents a greater number of false positive classifications than true detections using this logistic regression model.  However, it is also important to consider that accounts that have not been suspended could still be suspended in the future, or that some accounts in our data should be suspended but have succeeded in avoiding detection.

\begin{table} \center
\caption{Summary of sampled accounts from those incorrectly classified as suspensions using the regularized logistic regression model.}
\label{false-pos}
\begin{tabular}{lp{0.675\textwidth}}
\multicolumn{2}{c}{~}\\
\multicolumn{1}{c}{Screen Name} &
\multicolumn{1}{c}{Summary} \\ \hline
@abdulnagi313 & Few tweets, difficult to discern nature of account. \\[12pt] 
@445468a7e3fc45c & Very few tweets, user apparently follows ISIS activity and members on Twitter; possibly conducting research or surveillance.  \\[12pt] 
@613780 & Tweets Quranic verses in Arabic every few hours in consistent format; likely a Twitter bot. \\[12pt] 
@aarishmajeed & Account with no tweets following three ISIS-related media accounts. \\[12pt] 
@men9174 & Arabic-language pornography account followed by one of our seed accounts; following many other pornographic accounts. 
\end{tabular}
\end{table}

Sampling from the false positives resulting from this classification returned some accounts that were clearly ISIS supporters, supporting this notion that  many accounts should be or soon would be suspended.  Many of these ``false positives,'' however, were ISIS researchers, media, or otherwise difficult to discern.  Table \ref{false-pos} provides a summary of five randomly selected false positives found in our test data, when applying the classification probability threshold of 0.1.  The inclusion of the pornographic account @men9174 as a false positive is interesting and concerning.  Investigation reveals that this account is not following any of our ISIS seed accounts.   Our model classified this account with a probability of 0.101, very near our threshold, based primarily on its profile features.

\section{Detecting Multiple Accounts} \label{similarity}
Now that we have a model for predicting suspensions, the next question we address is whether we can automatically determine whether two accounts belong to the same user.  This question is relevant because we have observed many cases in which a user simply creates a new account after being suspended.  We have even found ISIS accounts dedicated to the purpose of broadcasting suspended users' new accounts to ISIS members and supporters.  By detecting multiple accounts belonging to the same
user, one can prevent extremist users from restarting their violative behaviors by creating new accounts and effectively
keep them suspended from the social network.

Twitter profiles essentially serve as avatars; the syntax and pictures provide cues about the identity of the account holder.  This is true for ISIS users as well and is intrinsic to the tactic behind the ISIS-based networks directed at recruitment.  As a result, when a suspended user opens a new account in Twitter, we have been able to identify it by comparing the names, images, screen names, and descriptions associated with each account.
This behavior is intuitive: these newly created accounts include cues to permit the suspended user's followers to identify and re-follow the recreated accounts.  We have found many examples of this predictable reiteration of account profile features in our data.  

\subsection{Suspended User Behavior}

In addition to regular account suspensions, we also observed that known ISIS users in our data set changed their screen names regularly.  We hypothesize that frequent screen name changes provide a means of avoiding tracking and detection, while retaining account information, friends and follower connections, and Twitter posts.  We also note that accounts that exhibit multiple screen name changes had higher suspension rates, which could mean that users are changing their screen names to avoid suspension.  

Table \ref{sally-sn} provides a timeline of screen name and name changes for two such accounts, purportedly belonging to British citizen Sally Jones who had adopted the online alias ``Umm Hussain al-Britani,' \cite{sally}.  Sally Jones and her husband, Junaid Hussain achieved celebrity status in ISIS, primarily due to Junaid Hussain's role in creating and leading the ``CyberCaliphate,'' as well as his previous involvement in the ``Team Poison'' hacking group.   Junaid Hussain was killed in a US airstrike in August 2015 \cite{junaid}. The timeline in Table \ref{sally-sn} was reconstructed from observed tweets, but the tweets from both of these accounts are no longer available due to account suspensions.  We note that the screen name changes became much more frequent when the user believed her behavior might result in suspension.  We also observe that in almost all cases, the user chooses some variant of the same online handle, e.g., ``OumHu55inBrit,'' which helps her retain her online identity and signals her status by announcing her attachment to Junaid Hussain, who always used the online alias ``Abu Hussain al-Britani'' (see follow-on discussion and Table \ref{comp-A}). 

\begin{table}[!hbt]
\centering \footnotesize
\caption{Partial screen name---tweet timelines for two Twitter user accounts purportedly belonging to Sally Jones.  These accounts have been suspended by Twitter and are no longer available.}\label{sally-sn}
\begin{tabular}{|l|c|}
\multicolumn{2}{c}{\bf First Account} \\ \hline
\multicolumn{1}{|c|}{\bf Tweet Time}
& \multicolumn{1}{c|}{\bf Screen Name}
\\ \hline
%

2015-09-30 11:45:37 
& OumHu554inBrit 
\\ \hline 
2015-09-30 19:58:15 
& OumHu554inBrit 
\\ \hline 
{\bf 2015-10-02 13:43:59} 
& {\bf \_Mrsl337} 
\\ \hline 
{\bf 2015-10-02 21:28:54} 
& {\bf OumHu554inBrit} 
\\ \hline 
{\bf 2015-10-03 00:48:01} 
& {\bf UmmHu55ain2} 
\\ \hline 
{\bf 2015-10-03 15:30:08}\dag 
& {\bf Oum1337} 
\\ \hline 
{\bf 2015-10-03 16:52:39} 
& {\bf OumHu554inBrit} 
\\ \hline 
2015-10-03 16:55:45 
& OumHu554inBrit 
\\ \hline 
2015-10-03 17:24:06 
& OumHu554inBrit 
\\ \hline 
{\bf 2015-10-03 23:31:29} 
& {\bf UmmHussain9ll} 
\\ \hline 
{\bf 2015-10-04 13:20:47} 
& {\bf OumHu554inBrit} 
\\ \hline 
\multicolumn{2}{c}{\bf Second Account} \\ \hline
\multicolumn{1}{|c|}{\bf Tweet Time}
& \multicolumn{1}{c|}{\bf Screen Name}
\\ \hline
{\bf 2015-10-05 16:44:55}\ddag 
& {\bf OumHu554in} 
\\ \hline 
2015-10-05 17:44:28 
& OumHu554in 
\\ \hline 
2015-10-05 20:36:22 
& OumHu554in 
\\ \hline 
{\bf 2015-10-06 18:03:26} 
& {\bf OumHussain} 
\\ \hline 
{\bf 2015-10-07 13:24:47} 
& {\bf OumHussa1n} 
\\ \hline 
\end{tabular}\\
 \begin{minipage}{0.85\textwidth}
\dag{}In this tweet the user warns she is about to release information that could get her suspended, and encourages her followers to be ready to retweet her. 
\\
\ddag{}This is the first tweet in a new user account, as the previous one was suspended. 
\end{minipage}
 \end{table}

While there might have been additional screen names and tweets associated with these accounts that we did not capture, we found the type of online behavior exhibited in Table \ref{sally-sn} indicative of many of the ISIS-supporting accounts in our data set.  Following suspension, the user apparently opens a new account and continues the same tactic, all the while adopting very similar account screen names and names.  Prominent ISIS members Sally Jones and Junaid Hussain provide examples of this behavior; accounts associated with them appear frequently in our ISIS data.  Querying our data for user accounts with a name similar to ``Umm Hussain Al-Britani'' returns 23 distinct entries, all of which have been suspended.

\begin{figure}
\centering
\includegraphics[width=3.5in]{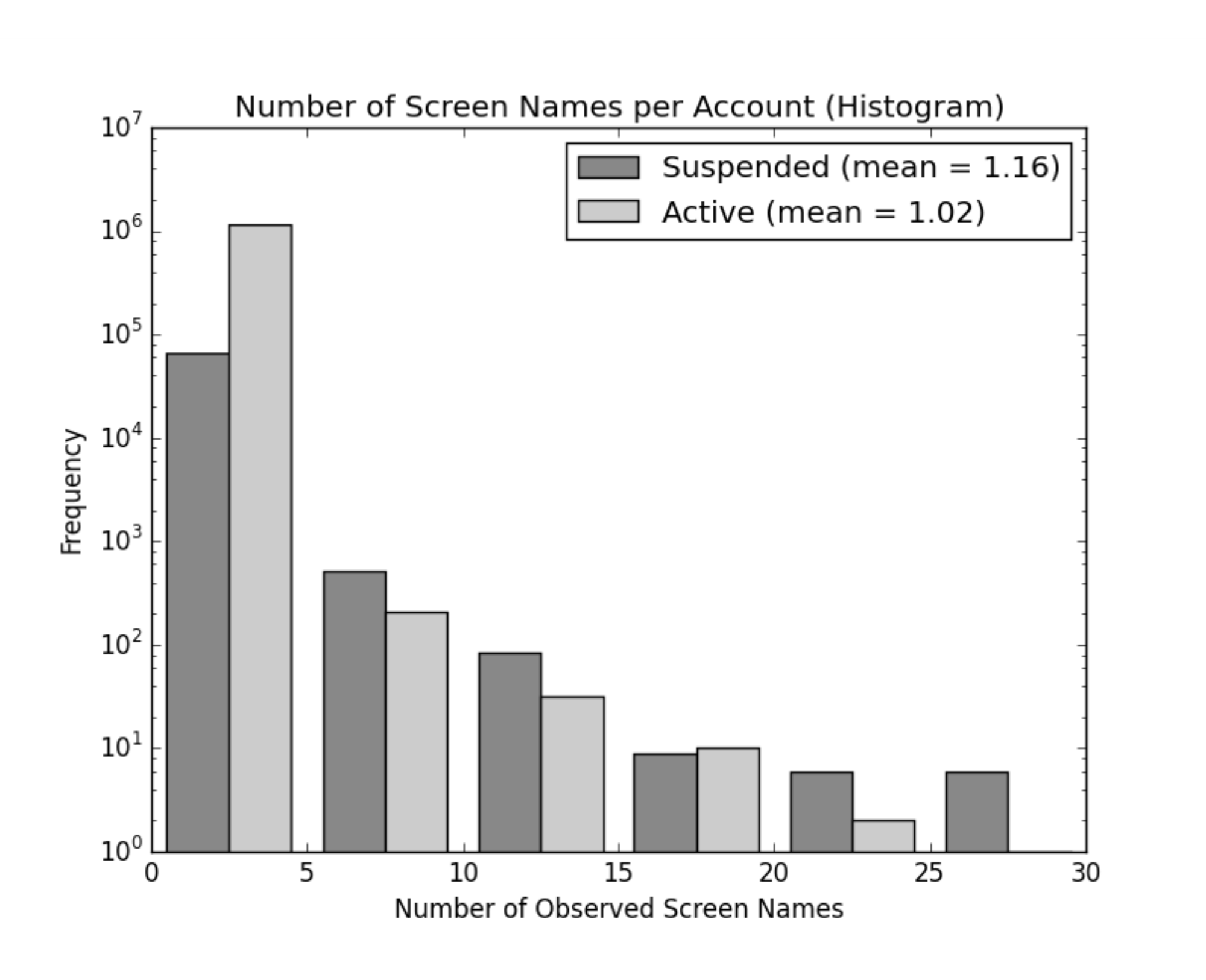}
\caption{Histogram of screen names for active and suspended accounts in our data.
The average numbers of screen names for suspended and active accounts are listed in the legend. } \label{sn-change-hist}
\end{figure}

Empirically, we found that we observed screen name changes in approximately 10\% of the accounts in our data that were eventually suspended, while in the accounts that remained active the number was close to 1\%.  Furthermore, anecdotal investigation of active accounts with multiple screen name changes suggests that many of these accounts are also ISIS-related.  
Figure \ref{sn-change-hist} provides a histogram comparison of the number of screen names associated with active and suspended accounts in our data set.  It is clear from the figure that the suspended accounts are much more likely to have more screen names.  For example, even though active accounts make up over 94\% of our data, only 18\% of the accounts with over 20 unique screen names are still active.  

These observations motivated our development of a method for locating new accounts belonging to a specific user. The first step in this process was to develop an automated method of identifying whether a pair of accounts belong to the same user.  To achieve this pairwise classification, we employ a supervised machine learning approach, which is described next. 

\subsection{Profile Comparison Metrics}

We define a Twitter user \emph{profile} as a vector of profile features $\v{x}$ associated with a Twitter account.  A Twitter account can only have a single user profile at any point in time.  The features of the profile are not fixed, however.  As we have noted, cases exist in our data in which users changed their screen name or other profile features, resulting in our obtaining multiple user profile feature vectors belonging to the same account.  

While it is possible for a single Twitter account to belong to different users at different times (e.g., an account gets hacked or one user simply provides the account login information to another), we assume that all of the profiles associated with the same Twitter account belong to a single user.  Our classification goal is therefore to compare two user profiles $(\v{x}^{(i)},\v{x}^{(j)})$ from \emph{different} Twitter accounts and identify whether or not they belong to the same user.

In order to train a model to perform this classification, we must construct profile comparison features from profile pairs $(\v{x}^{(i)},\v{x}^{(j)})$ that are useful in establishing whether they belong to the same user.  Building on our qualitative observations of individual ISIS Twitter users retaining identifying similarities between their multiple user profiles, we propose a set of similarity metrics based on comparisons of the following four profile features: screen name, user name,
profile picture, and profile banner image.
These similarity metrics are based on user profile characteristics that are publicly available on all accounts, even if the user has ``protected'' the account using Twitter's privacy settings. 

\subsubsection{Screen name and user name similarity metrics}

In comparing two screen names or two user names, we use the well-known Levenshtein ratio (see \cite{levenshtein}) to provide a measure of distance between two strings.  This ratio involves counting the number of character additions, deletions, or place exchanges required to transform one string into the other.  This number is normalized by the length of the longer string and then subtracted from one.  If we let $S$ be a set of strings of various lengths, the Levenshtein ratio can be thought of as a function $L: \; S^{2} \rightarrow [0,1]$ where $L(s_{1},s_{1})=1$ and $L(s_{1},s_{2})=L(s_{2},s_{1})$ for any $s \in S$.  $L(s_{1},s_{2})=0$ implies that strings $s_{1}$ and $s_{2}$ are not at all similar.

\begin{figure}[!hbt]
\centering
\includegraphics[width=3.25in]{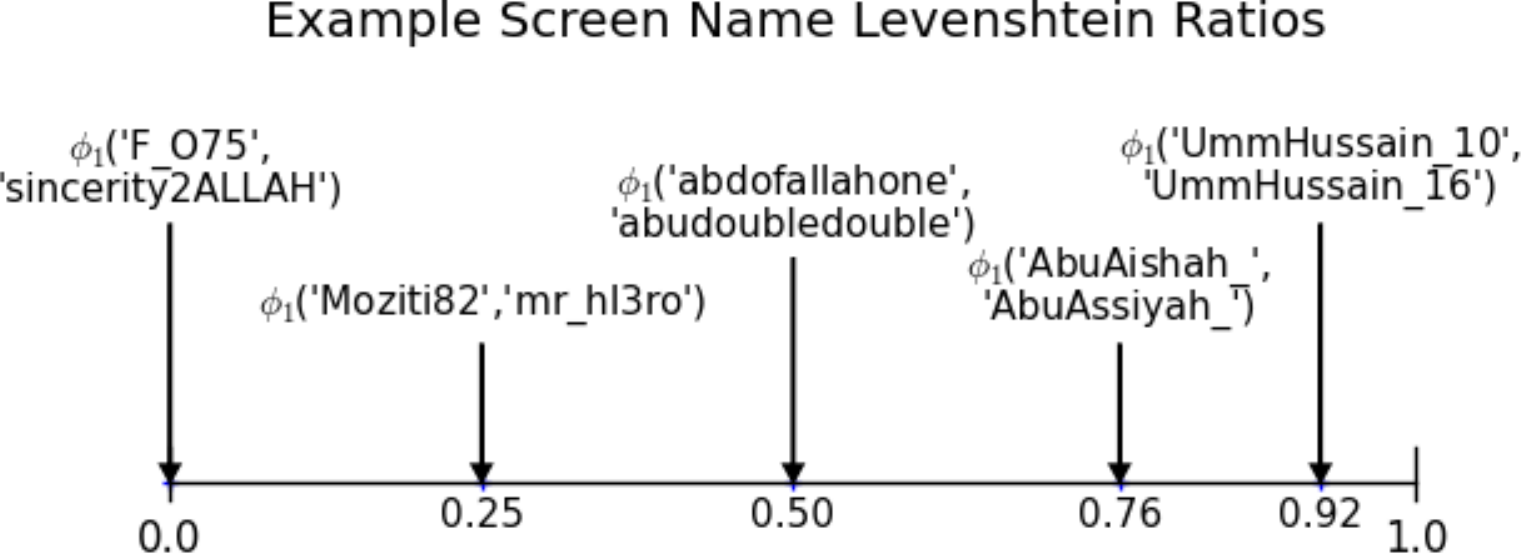}
\caption{Example screen name comparison Levenshtein ratios.} \label{SNLR-examples}
\end{figure}

Our first two comparison features, $\phi_{1}$ and $\phi_{2}$, are simply the screen name and user name Levenshtein ratios:
\begin{align*}
\phi_{1}(\v{x}^{(i)},\v{x}^{(j)}) = L(x_{SN}^{(i)},x_{SN}^{(j)}),
\phi_{2}(\v{x}^{(i)},\v{x}^{(j)}) = L(x_{N}^{(i)},x_{N}^{(j)}),
\end{align*}
where $x_{SN}^{(i)}$ and $x_{N}^{(i)}$ denote the respective screen name and user name of account profile $\v{x}^{(i)}$.
Figure \ref{SNLR-examples} provides an illustration of five screen name pairs and their corresponding Levenshtein ratios.

\subsubsection{Profile picture and profile banner similarity metrics}

We employ a simple image average hash algorithm (e.g., \cite{average-hash}) to compare two pictures.  Essentially, the algorithm partitions the image into $8\times 8$ equal-sized rectangular sub-images and then identifies whether the average shade of each sub-image is brighter or darker than the overall image average.  The algorithm runs efficiently and returns an $8\times 8$ binary matrix, which can easily be represented as a non-negative integer.

We denote the hash algorithm as a function $\v{H}: \Psi \rightarrow \mathbb{Z}_{+}$, where for any $\psi_{1},\psi_{2}\in \Psi$,
\begin{align*}
 \psi_{1}=\psi_{2} &\Rightarrow \v{H}(\psi_{1})=\v{H}(\psi_{2})\\
\v{H}(\psi_{1}) \neq \v{H}(\psi_{2}) &\Rightarrow \psi_{1} \neq \psi_{2} \\
\v{H}(\psi_{1})=\v{H}(\psi_{2}) & \Rightarrow \psi_{1} \approx \psi_{2}.
\end{align*}
Two images with the same hash value contain very similar patterns of shade.  Therefore, we assume that images with the same value are the same image.  
Our image similarity metric ($h$) is a simple step function that follows from this assumption:
\[
h: \Psi^{2} \rightarrow \{0,1\}, \quad h(\psi_{1},\psi_{2})=\begin{cases}
1 & \v{H}(\psi_{1})=\v{H}(\psi_{2}) \\
0 & \v{H}(\psi_{1})\neq \v{H}(\psi_{2}).
\end{cases}
\]

We use this image similarity metric to construct our third and fourth features:
\begin{align*}
&\phi_{3}(\v{x}^{(i)},\v{x}^{(j)})=h(x_{PP}^{(i)},x_{PP}^{(j)}) \\
&\phi_{4}(\v{x}^{(i)},\v{x}^{(j)})=h(x_{BP}^{(i)},x_{BP}^{(j)}), 
\end{align*}
where $x_{PP}^{(i)}$ and $x_{BP}^{(i)}$ are the respective profile and banner pictures for profile $\v{x}^{(i)}$.  These features are simply binary indicators for whether or not the images being compared have the same average hash matrix.

\subsection{Data Set Construction}

Having defined pairwise account profile similarity features, our next step was to clean the data and extract the features for use in a classification model.  Initially we examined 4,339 seed user accounts collected before June 4, 2015.  However, in order to keep the string similarity metrics consistent, we removed 395 accounts with user names strings that did not use the Latin alphabet.  This left us with 3,944 user profiles.  Within this set, we knew some user profiles we collected belonged to the same Twitter account and therefore the same user.  These accounts were identifiable by the Twitter user ID, which does not change even if a user changes his or her screen name or other profile features.  Our set of 3,944 profiles contained 3,855 unique Twitter accounts (i.e., unique user IDs), corresponding to 3,855 
 seed users.  For each pair of user profiles $(i,j)$, we computed a feature vector $\phi^{(i,j)}$ of the four similarity metrics.
This results in ${3,944 \choose 2}=7,775,596$ pairs.
%

\subsection{Data Labeling} \label{labeling}

We assume that each pair of user profiles either belong to the same user or belong to different users.  We denote this classification with binary class variable $y^{(i,j)}$, where
\[
y^{(i,j)}=\begin{cases}
1 & \mathrm{Profiles} \ i \ \mathrm{and} \ j \ \mathrm{belong \ to \ the \ same \ user} \\
0 & \mathrm{Profiles} \ i \ \mathrm{and} \ j \ \mathrm{belong \ to \ different \ users.}
\end{cases}
\]

Of the $7,775,596$ pairwise profile comparisons in our data, 95 could be traced to the same user because they actually belonged to the same account, identifiable by the Twitter user ID.  Although we do not seek to classify profiles belonging to the same account because we can already assume they belong to the same user, we left these comparison points in the data set as labeled data in order to train the classification model.  Updating profile features for an existing account is a different action than creating a new Twitter account, however, causing this labeled set to be biased toward profiles that are very similar.  
On the other hand, when a user creates a new Twitter account, he or she must deliberately set or leave blank each of the profile settings.  As a result, we do not expect the same level of similarity between two user profiles associated with separate accounts, but belonging to the same user, when compared to the similarity between two profiles belonging to the same Twitter account.

As a result, using these 95 pre-labeled data points for training might not be very useful for our purpose.  We also do not have any points classified as accounts belonging to different users.  To solve this problem, we labeled a subset of comparisons in our data set using the following method.
\begin{enumerate}
\item If profile $\v{x}^{(i)}$ and profile $\v{x}^{(j)}$ share the same user ID, set label $y^{(i,j)}=1$.  These are the 95 profile comparisons that are known to belong to the same user.
\item If profile $\v{x}^{(i)}$ and profile $\v{x}^{(j)}$ do not share the same user ID, \emph{and} 
\begin{equation}
(\v{x}^{(i)},\v{x}^{(j)}): 
\left\|\phi^{(i,j)}\right\|_{2}<0.1, \label{labelcrit2}
\end{equation}
we set label $y^{(i,j)}=0$.  These conditions establish that the profiles have very little in common, so we assume they belong to different users.  Table \ref{lowsim} provides an example of the features associated with a pair of accounts meeting this criterion.
\item Manually label a randomly selected subset of unlabeled pairs that exhibit relatively high similarity metrics.  We chose 168 pairs from the set of 1,257,350 pairs where
\begin{equation}
(\v{x}^{(i)},\v{x}^{(j)}): \left\|\phi^{(i,j)}\right\|_{2}>0.85, \label{labelcrit3}
\end{equation}
 for manual labeling.  In assigning a label to these pairs, we considered all available data in comparing the two profiles, including Twitter posting habits and account profile features, such as location and description, that are not considered in the model.    We found that 82 of these pairs were accounts belonging to the same user, while the remaining 86 of them were from different users.  Table \ref{highsim} provides an example comparison of the features of a pair of accounts meeting this criterion.
\end{enumerate}

\begin{table}[!htb]
\centering
\caption{Accounts exhibiting very low similarity, according to the selection criterion given in equation \eqref{labelcrit2}.}\label{lowsim}
{\footnotesize
\begin{tabular}{l|llc} 
Feature ($k$) & User $i$ & User $j$ & $\phi^{(i,j)}_{k}$ \\ \hline
User ID & 2683126250 & 3108319204 & [NA] \\  
Screen Name & khalidbinalwale & profomar0 & 0.08   \\ 
Name & Abu Muslim & prof & 0.00   \\ 
Profile Picture & 00\ldots{}c3 & 09\ldots{}cc & 0.00   \\ 
Profile Banner & 00\ldots{}00 & [None] & 0.00   \\[6pt]
 \multicolumn{4}{c}{$\|\phi^{(i,j)}\|_{2} = 0.08$}  \\ 
\end{tabular}
}
\end{table}

\begin{table}[!htb]
\centering
\caption{Accounts exhibiting very high similarity, according to selection criterion given in equation \eqref{labelcrit3}.  These accounts were manually labeled as belonging to the same user, i.e., $y^{(i,j)}=1$.} \label{highsim}
{\footnotesize
\begin{tabular}{l|llc} 
Feature ($k$) & User $i$ & User $j$ & $\phi^{(i,j)}_{k}$ \\ \hline
User ID & 3307258107 & 3297609231 & [NA] \\  
Screen Name & Ahmes\_Zirve\_\_ & Ahmes\_\_Zirve & 0.88   \\ 
Name & Ahmes Zirve & Ahmes Zirve & 1.00   \\ 
Profile Picture & ff\ldots{}ff & ff\ldots{}ff & 1.00   \\ 
Profile Banner & [None] & [None] & 1.00   \\[6pt]
\multicolumn{4}{c}{$\|\phi^{(i,j)}\|_{2}=1.94$}  \\ 
\end{tabular}
}
\end{table}

\subsection{Classification Model}

From our set of labeled data, we set aside 10\% for out of sample evaluation of model performance.  This percentage was enforced for each of the three labeling methods, so that the test set included 10\% of the hand labeled data points, for example.  We then fit an $L_{1}$-regularized logistic regression model on the training data.  In other words, we assume
\[
\mathbb{P}(y^{(i,j)}=1)=\left(1+e^{\beta^{T}\phi^{(i,j)}+\beta_{0}}\right)^{-1}
\]

We identified $\lambda=10$ as the regularization parameter that provided the best performance in cross validation.  The intercept and coefficients for the logistic regression model fit on the training data are shown in Table \ref{table:regression_matching}.  Interestingly, profile banner similarity is not useful in this model in determining the probability of two profiles belonging to the same user.

\begin{table}[!htb]
\centering
\caption{Regression coefficients for matching accounts.}\label{table:regression_matching}

{\footnotesize
\begin{tabular}{|c|c|} 
\hline
Feature & Regression coefficient \\ \hline
Intercept & -8.05\\ \hline
Screen name Levenshtein ratio ($\phi_1$) & 2.94\\ \hline  
User name Levenshtein ratio ($\phi_2$) & 7.05\\   \hline
Profile picture hash matrix ($\phi_3$) & 1.88\\   \hline
Banner picture hash matrix ($\phi_4$) & 0\\   \hline
\end{tabular}
}
\end{table}

\begin{figure}[!htb]
\centering
\includegraphics[scale=0.5]{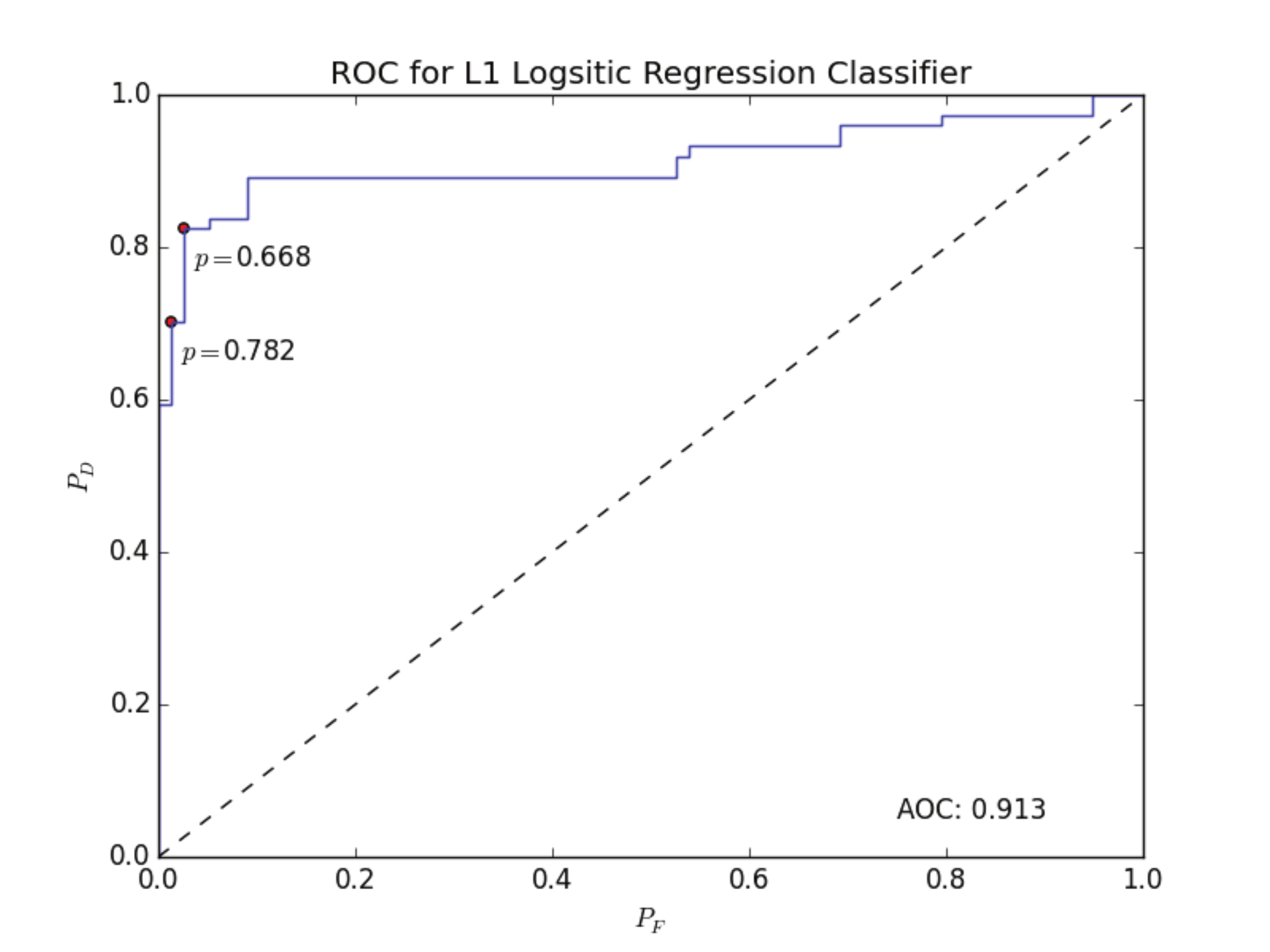}
\caption{Logistic regression ROC curve on hand labeled data.} \label{L1ROC}
\end{figure}

The receiver-operator characteristic (ROC) curve plotted \emph{for the manually labeled training and test data combined} is given in Figure \ref{L1ROC}.  ROC curves plotted separately for the training and test data were very similar, and classification on the test data points that were not manually labeled (i.e., they were labeled using steps (1) or (2) of the labeling method given in section \ref{labeling}) was nearly perfect.  The AUC in Figure \ref{L1ROC} is approximately 0.91.  

We view the ROC curve on the manually labeled data in Figure \ref{L1ROC} as an approximation for the ``worst case'' performance of the classifier.  We selected these pairs for manual labeling because they exhibited some degree of similarity, based on the $L_{2}$ norm of the comparison feature vector, anticipating that they would be among the most difficult points to classify.  As noted previously, plotting the ROC curve on all of the labeled data, or on the entire test set, shows near perfect classification.

Because we anticipate that most account pairs belong to different users, maintaining a low false positive misclassification rate is important. A small false positive rate could equate to a large number of misclassified points.  For this reason, we select a false positive threshold of 2\% on the hand-labeled ROC curve.  This threshold leads us to a classification probability threshold of 0.782, as indicated in Figure \ref{L1ROC}.  In other words, we assign a classification $\hat{y}^{(i,j)}$ to a profile pair $(\v{x}^{(i)},\v{x}^{(j)})$ according to the function
\begin{equation}
\hat{y}^{(i,j)} = \begin{cases}
1 & \left(1+e^{\beta^{T}\phi^{(i,j)}+\beta_{0}}\right)^{-1} \geq 0.782 \\
0 &  \left(1+e^{\beta^{T}\phi^{(i,j)}+\beta_{0}}\right)^{-1} < 0.782.
\end{cases} \label{classifier}
\end{equation}

Based only on the hand-labeled data ROC, we expect this classifier to correctly identify over 80\% of account pairs belonging to the same user  while misclassifying less than 2\% of the account pairs belonging to different users.  Because the manually labeled data consists of account pairs that exhibit some substantial measure of similarity, we expect performance on the entire data set to be much better, similar to the near-perfect classification on the test data.

\begin{figure}
\centering
\fbox{
\includegraphics[scale=0.25]{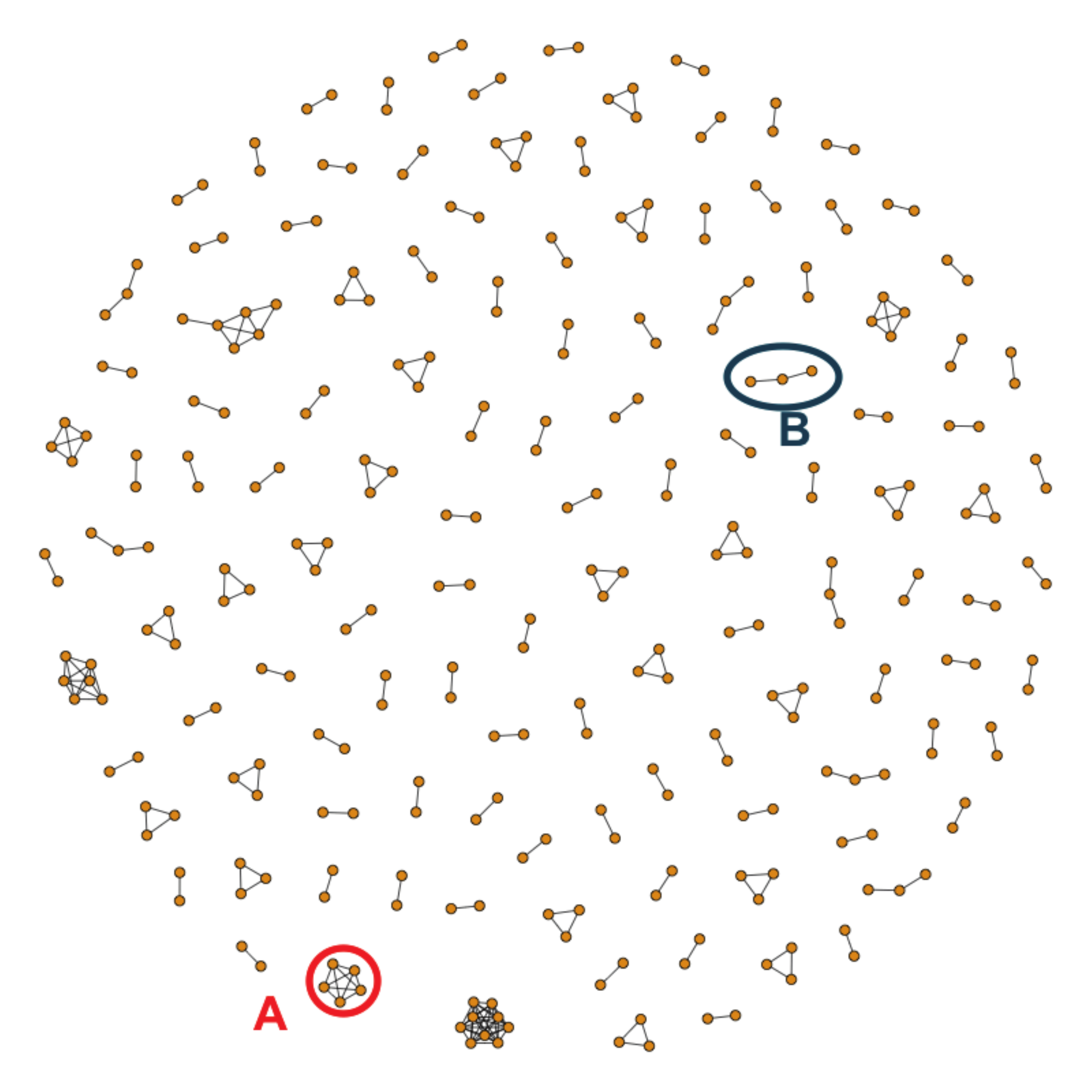}
}
\caption{Graph representation of accounts belonging to the same user using our regression model and equation \eqref{classifier} with a threshold of 0.782.} \label{merge-graph}
\end{figure}

When we apply the classifier in equation \eqref{classifier} to the entire data set, we obtain 318 account pairs classified as belonging to the same user.  Sixty-two of these pairs have the same account ID and are therefore known to be from the same account, while the remaining 256 pairs come from different Twitter accounts.  Figure \ref{merge-graph} provides a network representation of these account connections.  Each node in the plot represents a unique Twitter account.  An edge drawn between two accounts indicates our classification equation labels the pair of accounts as belonging to the same user.  Only accounts with at least one edge are depicted in Figure \ref{merge-graph}.

Most of the components in the graph depicted in Figure \ref{merge-graph} are fully connected, which is as we would expect.  Component A is an example of a fully connected component, consisting of five Twitter accounts.  These account profile features are listed in Table \ref{comp-A}.  They are all very similar and indeed appear to belong to the same user.

Component B, on the other hand, consists of three accounts but is not fully connected.  Table \ref{comp-B} provides  a list of the profile features associated with these three accounts.  While they all appear to belong to the same user, comparison of the first and third profiles given in the table resulted in probability
$
\mathbb{P}(y^{(i,j)}=1)=0.774,
$
which falls below our classification threshold.  While in this case these two accounts are connected by way of a third account that meets the classification threshold with both of them, it is clear that setting threshold this high does indeed miss some pairs of accounts that probably do belong to the same user.  We discuss the sensitivity of the results as a function of the classification threshold in Appendix \ref{app:sensitivity}.

\begin{table} \centering
\caption{Accounts comprising component A.  While average hash values for profile pictures are abbreviated, they are the same for all profiles.} \label{comp-A}
\begin{tabular}{|l|l|l|} 
\multicolumn{3}{c}{} \\ \hline
 Screen Name & Name & Profile Pic \\ \hline
AlJabarti28 & Abu Yusuf Al-Jabarti & 20\ldots{}00\\ 
BanuKombe & Abu Yusuf Al-Jabarti & 20\ldots{}00\\ 
enkorela & Abu Yusuf Al-Jabarti & 20\ldots{}00\\ 
ouaicheu & Abu Yusuf Al-Jabarti & 20\ldots{}00\\ 
ouaisheu & Abu Yusuf Al-Jabarti & 20\ldots{}00\\ \hline
\end{tabular}
\end{table}

\begin{table} \centering
\caption{Accounts comprising component B.} \label{comp-B}
\begin{tabular}{|l|l|l|} 
\multicolumn{3}{c}{} \\ \hline
Screen Name & Name & Profile Pic \\ \hline
Aqidahhaqq & Colonel Shaami & [None]\\ 
AnsarAlUmmah49 & Colonel Shaami & [None]\\ 
buruan8 & Colonel Shaami & [None]\\ \hline
\end{tabular}
\end{table}

\section{Refollowing Model} \label{refollowing}
In the previous section we used machine learning to produce a  method for efficiently finding groups of accounts that are likely to belong to a single user. In this section, we use the account clusters produced from this method in an effort to learn how users tend to reconnect, or \emph{refollow}, other user accounts when opening a new account. 

Suppose a user $t$ has his account suspended and decides to open a new account.  After getting the account open, $t$ decides to follow some other users.  We have observed that in many cases, $t$ will refollow at least some of the user accounts he was previously following with his suspended account, and it seems reasonable to assume that any suspended user would want to reconnect with some of the same people he or she was following prior to suspension.  

In this section we fit a probability model that assigns a value to each of $t$'s former friends, giving the probability $t$ will refollow the former friend upon opening a new Twitter account.  We again turn to logistic regression as a means to producing this probability model.

\subsection{Data}

Using the logistic regression model from Section \ref{similarity} with a cutoff of 0.782, we grouped the seed accounts into clusters, each of which we assume belong to the same user.  A network representation of the non-singleton clusters is shown in Figure \ref{merge-graph}.  Accounts in each cluster were then sorted by account age.  After sorting, we compared the friend lists of each pair of consecutive accounts.  For each friend  of the former account, we created a row in our data set labeled with an indicator of whether or not the same friend was connected to the latter account.

\begin{table}[!hbt]
\centering
\caption{Example of @MusabGharieb18's (@M\ldots18) refollowing behavior upon opening new account @MusabGharieb13  (@M\ldots13).} \label{acctcomparison}
\begin{tabular}{l|cc}
Friend & @M\ldots18 & @ M\ldots13 \\ \hline
@poorslave\_3 & YES & YES \\
@enkorela & YES & NO \\
@StillUkhtMaryam & YES & YES\\
@Yaqub\_London & YES & NO \\
\end{tabular}
\end{table}

For example, user account \#3280844606 (@MusabGharieb18) and user account \#3343999888 \mbox{(@MusabGharieb13)} are consecutive accounts belonging to the same user cluster.  Table \ref{acctcomparison} shows whether each account was following certain friend accounts.  Table \ref{datarows} shows how each of @MusabGharieb18's friends would then generate a row in the data for this logistic regression model.

\begin{table}[!hbt]
\centering
\caption{Example data rows resulting from refollowing behavior given in Table \ref{acctcomparison}.  Features are omitted but include, for example, characteristics from each friend's profile.} \label{datarows}
\begin{tabular}{l|cc}
Friend & Features & \parbox{1in}{\centering Refollowed \newline (Response)} \\ \hline
@poorslave\_3 & $\cdots$ & 1 \\
@enkorela &$\cdots$ & -1 \\
@StillUkhtMaryam & $\cdots$ & 1\\
@Yaqub\_London &$\cdots$ & -1 \\
\end{tabular}
\end{table}

\subsection{Features}

In order to obtain a good fit, we included features from the suspended user's earlier account (e.g., @MusabGharieb18) as well as features from the friend account (e.g., @poorslave).  For a suspended user account {\bf User0} that was following account {\bf Friend}, we construct a variety of features which can be broken down into different categories.  One set of features
deals with the features of the individual accounts of  {\bf User0} and {\bf Friend}.  A related set of features are about the similarity of the two accounts.  There is a category of features that deals with the
interactions between  the two accounts.  Finally, there is a category of features that describe aggregate properties of the neighbors of {\bf User0}.  A complete list of the features used in our model can be found in Appendix \ref{app:features_list}.

\subsection{Kernel Logistic Regression}

Intuitively, some interactions among our set of features might be more predictive than the features themselves.  For example, the average number of {\bf User0}'s friends might not be very useful in estimating the probability {\bf User0} refollows a specific {\bf Friend} account.  However, this value multiplied by {\bf Friend}'s number of Twitter friends could be very useful.  For this reason, we use a quadratic kernel in this logistic regression model, which ensures the regression is fit on all linear and quadratic terms, including pairwise interactions:
\[
K(\vx,\mathbf{y}) = (1+\vx^{T}\mathbf{y})^{2}
\]

Given a training data set $\{\vx_{1},\vx_{2},\ldots,\vx_{N}\}$, the corresponding logistic regression model is
\[
\hat{p}(\vx) = \left(1+e^{\sum_{i=1}^{N} \alpha_{i}K(\vx,\vx_{i})}\right)^{-1}.
\]

The parameters $\mathbf{\alpha}=(\alpha_{1},\ldots,\alpha{N})$ are fit on the training data using an $ L_{2}$-regularized log loss:
\[
\mathbf{\alpha} = \arg\min_{\hat{\mathbf{\alpha}}}
\sum_{i=1}^{N} \log (1+e^{-y_{i} \sum_{i'=1}^{N}\hat{\alpha}_{i} K(\vx_{i},\vx_{i'})}) + \lambda \hat{\mathbf{\alpha}}^{T}\hat{\mathbf{\alpha}},
\]
where $y_{i}$ is the response in the $i$th row of the training data.  These responses take value -1 if the {\bf Friend} was not refollowed, or 1 if the {\bf Friend} was refollowed, as annotated in Table \ref{datarows}.  The parameter $\lambda$ serves as the regularization coefficient.

\subsection{Performance}

\begin{figure}[!htb]
\centering
\includegraphics[scale=.33]{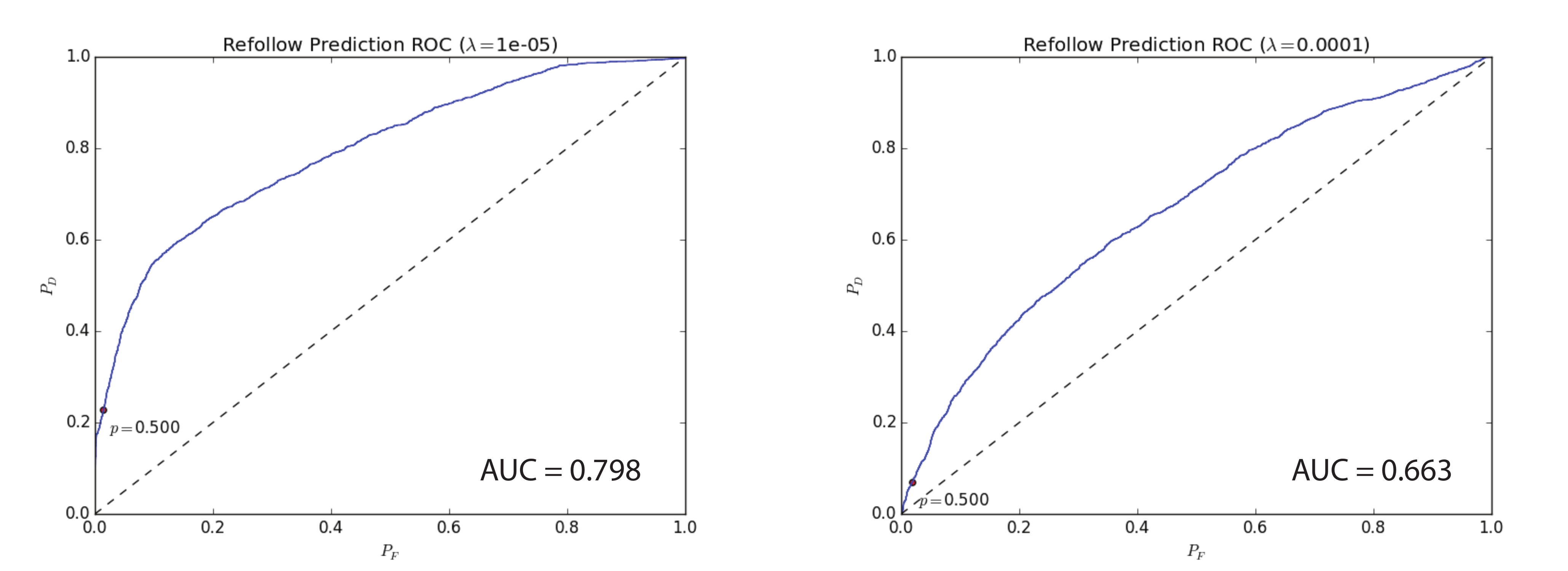}
\caption{ROC curve for $L_{2}$-regularized quadratic kernel logistic regression performance on out-of-sample test data. (left) Test data
and training and validation data can contain the same user.
(right) Test data and training and validation data do not contain the same users.} \label{refriend-ROC}
\end{figure}

In order to fit this model we used gradient-based optimization methods available in Python's {\tt scipy} package \cite{scipy}.  We first selected training (50\%), validation (25\%), and test (25\%) sets randomly from all of the rows of the data and normalized the entire data set based on the values in the training data.  Through validation we found that $\lambda=10^{-5}$ provided the highest AUC.  Performance on out-of-sample test data is depicted in Figure \ref{refriend-ROC}.

From the figure it appears that we can predict with some accuracy which former friends a suspended user is likely to reconnect with.  It is possible, however, that the model is learning refollowing preferences of individual users in the data set.  To investigate this possibility, we selected new training, validation, and test sets by randomly selecting different \emph{user clusters} for each and included all of the rows corresponding to these user clusters in the corresponding set. In other words, each \emph{component} depicted in Figure \ref{merge-graph} was assigned as a whole to either the training, validation, or test data, approximately maintaining the 50\%-25\%-25\% ratios.  Unlike the previous data partition, this constraint would ensure that all of the rows in Table \ref{datarows} went to the same set, because they belong to the same user.

Using this new data partition, validation and testing were completed on data consisting of entirely different users than those that provided the training data.   Through validation we found that $\lambda=10^{-4}$ provided the highest AUC on this new data partition.  Out-of-sample performance suffered, as can be seen in the ROC plot in Figure \ref{refriend-ROC}.  Comparing the performance on each partition provides some interesting insights.  First, the AUC for the new partition in Figure \ref{refriend-ROC} is 0.66, which indicates that there is some underlying refollowing behavior that transcends the users in our data.  However, our ability to predict whether or not a suspended user will refollow an old friend increases substantially when we include that user's past behavior in the training data.  The difference in performance gives us an idea  of how useful it is to have data on a specific user's past behavior when predicting whom the user will refollow.

Because we used a quadratic kernal logistic regression, the expressions for the fit models are not easy to interpret.  Their performance shows that we can predict with some accuracy the refollowing behavior of a suspended user, even in the absence of previous refollowing behavior, based solely on the refollowing behavior of others.  We make use of this capability in the next section, where we develop a method to search for a suspended user's new account.  In practice an analyst might be able to produce a much better model for a specific user by carefully incorporating past refollowing behavior, if available.

\section{Suspended User Search} \label{search}

We now make use of our findings from the previous sections to address another relevant problem.  We have observed multiple incidences in our data of suspended users quickly creating a new Twitter account in order to continue their unethical activity, as exemplified in Table \ref{sally-sn}.  In these instances it would be useful for those tasked with monitoring nefarious users, such as social media service providers or intelligence community personnel, to find an efficient way to search for the suspended user's new account.  

We assume we are given a \emph{target} user whose account has been suspended by Twitter.  We have stored the target user's account information, including lists of the target's friends and followers.  From this information we wish to locate the target user's new Twitter account, if one exists, as efficiently as possible.  Our approach to solving this problem is to query the followers of each of the target user's known Twitter ``friend'' accounts, prior to suspension, and search the results for a new account belonging to the target user.

\begin{figure}[!htb]
\centering
\fbox%
{
	\includegraphics[width=0.4\textwidth]{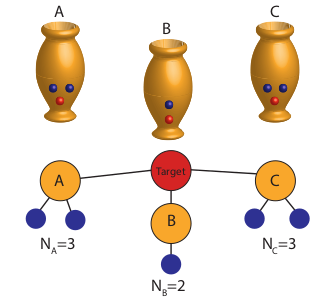}
}
\caption{Network search representation as a multi-urn model.} \label{fig:urn_network}
\end{figure}

Our network search model builds directly on the multi-urn search model presented in \cite{urn1} and is illustrated in Figure \ref{fig:urn_network}.  We can think of each of the target's former friends $i$ as an urn containing $N_{i}$ marbles, which represent the neighbors of $i$.  If the target has connected to former friend $i$, then he is among $i$'s neighbors and a single red marble is one of the $N_{i}$ marbles in urn $i$.  Excepting these red marbles, all marbles in all urns are blue.

Each follower query can be thought of as choosing a nonempty urn $j$ in the multi-urn model and removing some fixed number of its marbles.  The number of marbles removed is determined by the query method used and, unlike the search model in \cite{urn1}, can be more than one.  Having a red marble among those removed represents finding the target user's account, and the search terminates.  

\subsection{Suspended User Search Model}

Let $\mathcal{V}$ be the set of \emph{known friend} accounts.  These are the accounts that the target user was following prior to begin suspended.  For each known friend $i \in \mathcal{V}$, let $N_{i}$ be the number of Twitter accounts that are following $i$.  These quantities are easily obtained through the Twitter API.  

Using the Twitter API it is possible to obtain a list of the followers of a specified user, provided the user has not enabled privacy protection on the account.  Twitter offers two methods for executing these queries: {\tt GET followers/list} and {\tt GET followers/ids}.  Both methods are rate limited to 15 queries within any 15-minute time period.  {\tt GET followers/list} returns standard Twitter user profile information for each follower, but only returns up to 200 profiles per query.  {\tt GET followers/ids} has the same rate limit, but returns up to 5,000 user IDs per query \cite{rest}.

Each method can be cursored so that subsequent queries of the same user continue to produce unique results, until all of the user's followers' profiles or IDs have been obtained.  For our analysis, we set $N_{M}$ as the maximum number of unique followers obtained per query, although in practice we assume this number to be 5,000 as established in the {\tt GET followers/ids} method.  Therefore if we have queried user $i$'s followers $n$ times, we expect the next query of user $i$'s followers to return
$\min\{N_{M},N_{i}-nN_{M}\}$ new results, provided $i$ still has unqueried followers ($N_{i}-nN_{M}>0$).

Additionally, we make the following assumptions:
\begin{enumerate}
	\item After being suspended, the target user creates a new account with probability $\rho_{0}$, which we refer to as the a priori \emph{existence probability}.  If the target has not created a new account, then he does not have a node in the network and will not be found through follower queries.  The value of $\rho_{0}$ quantifies the searcher's belief that the target exists in the network.
	\item If the target user creates a new account, he reconnects with each former friend $i \in \mathcal{V}$ with some probability $\varphi_{i}$, which can be estimated from previous account data as was done in Section \ref{refollowing}. We refer to this as the \emph{reconnection probability} to former friend $i$.
	\item Reconnections to former friends are independent; whether or not the search target reconnects with former friend $i$ does not affect the probability he reconnects with former friend $j \neq i$.
	\item If the target user is following user $i \in \mathcal{V}$, then each account returned in each query of $i$'s followers is equally likely to be the target's account.  
	\item The searcher can quickly and accurately determine whether an account obtained from a follower query is the target user's account.  This can be done using the approach developed in Section \ref{similarity}.
\end{enumerate}

The search process is modeled as the execution of follower queries in discrete stages.  In each stage $t \in 0,1,\ldots, N-1$, the searcher chooses one of the target user's former friend accounts and executes a follower query.  Here, $N$ is the total number of queries required to examine all of the followers of all former friends, and is assumed to be finite.  If the target user's new account is among the query results, the search terminates.  Otherwise, the searcher executes another query unless all $N$ queries have been exhausted or the searcher concludes that the target has not created a new account.


The objective of the search is to minimize the total number of queries.  In order to remain consistent with the multi-urn search model in \cite{urn1}, we do not consider the cost of a query that succeeds in returning the target user's new account.  Therefore, the objective in our search model is to minimize the number of \emph{unsuccessful} search queries.  The best result possible would be to find the search target in the first query, in which case there are zero unsuccessful queries.  Because of the stochastic nature of this process, we say that a search policy is optimal if it minimizes the \emph{expected} number of unsuccessful queries.


\subsection{Initialization}

We assume that data collected on the target user provides a list of known former friend accounts.   Using the Twitter API, it is relatively easy to determine which of these accounts are still active, whether or not they are ``private,'' and their follower counts.  We initialize set $\mathcal{V}$ as the set of all former friend accounts that are active at the time of search execution, that have followers that can be queried (i.e., have a positive number of followers and are not ``private'' accounts).   We use the follower counts for these accounts to initialize $N_{i}, \ i \in \mathcal{V}$.


This search model also requires an initial probability that the target user would reconnect with each former friend $i \in \mathcal{V}$, given he has created a new account.  Let $A$ be the event that the target has created a new account, $B_{i}$ be the event that the target is following former friend $i \in \mathcal{V}$, and $B=\bigcup_{i\in \mathcal{V}}B_{i}$ be the event that the target has reconnected with at least one former friend.  From our definitions above, we can write
$
\varphi_{i} = \P(B_{i}|A).
$
We can obtain the value of this probability using the approach presented in Section \ref{refollowing}.
Note that event $B$ can also be interpreted as the event we can find the target user by exhaustively querying the followers of all former friends.  Using our independence assumption we have
\[
\P(B^{c}|A)= 1- \P(B|A)= \prod_{i\in \mathcal{V}} (1-\varphi_{i}).
\]

We also must select a value for the a priori existence probability $\rho_0$,which can be done
based on the beliefs of experts in the relevant domain.  As the search process progresses,
the conditional existence probability will evolve.  
The search terminates if the target user is found or all follower queries are exhausted.  In addition to these criteria, a searcher might want to terminate the search upon achieving reasonable certainty that the target user has not created a new account.  We allow for this termination criterion by including a termination conditional existence probability $\bar{\rho}$.  If at any stage the conditional existence probability falls below $\bar{\rho}$, the search terminates and the searcher concludes that the target has not created an identifiable new account.  If $\bar{\rho}$ is set to zero, then the search continues until the target is found or until all follower queries are exhausted.

\subsection{The Discrete Stochastic Search Process}

As we have suggested, the search process can be modeled as a set of urns, each representing a former friend.  Urn $i \in \mathcal{V}$ has $N_{i}$ marbles, which represent former friend $i$'s followers.  In each stage the searcher chooses a former friend (or urn) and executes a follower query, receiving up to $N_{M}$ results (or drawing at most $N_{M}$ marbles from the urn).  The search continues until one of the following occurs:
\begin{itemize}
\item The target's new account is found (a red marble is among those drawn),
\item  The probability the target has created a new account falls below the termination probability $\bar{\rho}$
\item The queries of former friends' followers are exhausted (there are no marbles left in any of the urns).
\end{itemize}

\subsubsection{Policy}

Suppose we consider a valid policy as any sequence of former friend queries in which each former friend is exactly exhaustively queried.  In other words, if we let $\vu=(u_{0},u_{1},\ldots,u_{N-1})$ be a policy in which former friend $u_{t}\in \mathcal{V}$ is queried in stage $t$, then $\vu$ is valid if and only if 
\[
|\{t 
: u_{t}=i\}|=\left\lceil \frac{N_{i}}{N_{M}}\right\rceil 
 \quad \forall \; i \in \mathcal{V}.
\]

Notice that any valid policy can be completely specified in advance as an ordering of follower queries that is executed until one of the three termination criteria are met.  As long as the target is not found, state transitions are deterministic and can be enumerated a priori.  Except for the decision to terminate, there is no benefit to making policy decisions during the search.  Unsuccessful search results do not provide any additional insight into which ordering of queries might yield a lower cost.

\subsubsection{System State and Transitions}

In order to analyze the dynamics of the system we define the system \emph{state}, $\vx(t)$, at stage $t$ as either 
a $|\mathcal{V}|$-dimensional vector in which the $i$th element $x_{i}(t)$ is the number of follower queries that have been executed on former friend $i \in \mathcal{V}$ in previous stages, or
a terminal state, ``Terminate.''
At stage $t=0$, no queries have been executed and presumably the search has not terminated, so that $\vx(0)=\mathbf{0}$.  In any non-terminal state, let the vector $\vx(t)$ be specified as a function of the policy being executed:
\begin{equation}
x_{i}(t) = |\{\ell < t: u_{\ell}=i\}| \quad \forall \; i \in \mathcal{V}. \label{eq: u_to_x}
\end{equation}


State transitions in this system are a function of the current state, the policy, and a stochastic input representing whether the target account is found as a result of the current stage query.  Let
\[
w(\vx(t),i) = 
\begin{cases}
0, & \text{Target is not found querying } i \text{ from state } \vx(t)\\
1, & \text{Target is found querying } i \text{ from state } \vx(t).
\end{cases}
\]

We now have all of the definitions needed to write the state transition function that governs this search model.
\begin{align*}
\vx(t+1)&=f(\vx(t),u_{t},w(\vx(t),u_{t})) 
\\
&= 
\begin{cases}
\mathrm{``Terminate,''} & w(\vx(t),u_{t})=1 \ \text{or other termination criterion are met} \\
\vx(t) + \mathbf{e}_{u_{t}}, & \text{otherwise.} \\
\end{cases}
\end{align*}
Here, $\mathbf{e}_{i}$ represents the $i$th unit vector.

\subsection{Search Process Dynamics}

We define the function 
\begin{align*}
\psi_{i}(\vu,t)&= \max\left\{\frac{x_{i}(t)N_{M}}{N_{i}},1\right\}
\\
&= \begin{cases}
\frac{x_{i}(t)N_{M}}{N_{i}} & x_{i}(t) = 0,1,\ldots,\left(\left\lceil\frac{N_{i}}{N_{M}}\right\rceil-1\right)\\
1 & x_{i}(t) = \left\lceil\frac{N_{i}}{N_{M}}\right\rceil
\end{cases}
\end{align*}
as the fraction of former friend $i$'s followers that have been queried before stage $t$ when executing valid policy $\vu$ (or, using the urn analogy, the fraction of marbles that have been removed from urn $i$ at stage $t$), conditioned on not having found the target user prior to stage $t$.  This function captures the assumption that, provided former friend $i$ has more than $N_{M}$ unqueried followers remaining in stage $t$, the query returns $N_{M}$ followers.  If former friend $i$ has fewer than $N_{M}$ unqueried followers remaining in stage $t$, then the query will return all of the remaining unqueried followers.  

This function is  strictly increasing at a constant rate of $\frac{N_{M}}{N_{i}}$ as $x_{i}(t)$ increases from 0 to $\left\lceil \frac{N_{i}}{N_{M}} \right\rceil-1$.  It continues to increase, at a possibly slower rate, in the $\left(\lceil \frac{N_{i}}{N_{M}} \rceil\right)$th query of former friend $i$.  Because $x_{i}(t)$ is nondecreasing in $t$, we can conclude that $\psi_{i}(\vu,t)$ is also nondecreasing in $t$.

For example, suppose a certain former friend has 12,000 followers and that each follower query returns at most $N_{M}=5,000$ followers.  Then, the first and second query of this former friend will return 5,000 followers each, while the final query will only return 2,000 followers.  In general, we expect the first $\left\lceil \frac{N_{i}}{N_{M}}\right\rceil-1$ follower queries of former friend $i \in \mathcal{V}$ to return $N_{M}$ results, while the final query returns $N_{i} - 
\left(\left\lceil\frac{N_{i}}{N_{M}}\right\rceil-1\right)N_{M}$ results.  This irregularity results in final queries of former friends to affect the system dynamics differently than the preceding queries of the same former friends.

\subsubsection{Conditional Existence Probability}
We now develop an expression for the conditional existence probability, i.e., the probability that the target user has created a new account conditioned on having reached a certain non-terminal state, $\vx(t)$.  Let $A$ be the event that the target user has created a new account.  For simplicity of notation, we condition directly on the state vector $\vx(t)$ to denote the event that this state has been reached without finding a target user's new account, so that
$
\rho(t) = \P(A|\vu,\vx(t))
$
is the new account existence probability conditioned on having reached state $\vx(t)$ when executing valid search policy $\vu$ without having found the target account.  Note that $\rho(0)=\rho_{0},$ the initialization value.

Using Bayes' rule, the conditional existence probability is 
\[
\rho(t) = \rho_{0}
\left(
\frac{
	\prod_{
		i \in \mathcal{V}
	}
	\left(
		1-\psi_{i}(\vu,t)\varphi_{i}
	\right)
}
{
	1-\rho_{0}+\rho_{0}\prod_{
		i \in \mathcal{V}
	}
	\left(
		1-\psi_{i}(\vu,t)\varphi_{i}
	\right)
}
\right).
\]
The terms inside the products are the probabilities of not finding the target account among the followers of each former friend $i$, given that $\psi_{i}(\vu,t)$ of those followers have been queried and examined.  Multiplying these probabilities together implicitly relies on our assumption that the target user reconnects to his former friends independently.

The expression for $\rho(t)$ is the initial existence probability multiplied by a ratio of two linear functions of the product $\prod_{i \in \mathcal{V}}(1-\psi_{i}(\vu,t)\varphi_{i})$.  Because $\psi_{i}(\vu,t) \leq 1 \ \forall i \in \mathcal{V}$ and is nondecreasing in $t$, $\prod_{i \in \mathcal{V}}(1-\psi_{i}(\vu,t)\varphi_{i})$ is nonincreasing in $t$.  The coefficient in the denominator ($\rho_{0}$) is no more than that of the numerator ($1$), and therefore the conditional existence probability is nonincreasing in $t$ and converges to 0 as $\prod_{i \in \mathcal{V}}(1-\psi_{i}(\vu,t)\varphi_{i})$ decreases to 0. 
This monotonicity property aligns with intuition: the more the social network is searched without finding the target user, the less likely it is that the target user exists in the network.  

Other than the conditional existence probability at each stage, the value of the initial existence probability $\rho_{0}$ does not affect the system dynamics.  Implicit in the execution of the search is the assumption that the search target has created a new account and reconnected to former friends in a way that can be represented by a probability model.  The utility of including an existence probability in the model is that it enables the searcher set a search termination criterion when he is sufficiently convinced that the target user has not created a new account, based on the value of the conditional existence probability.


\subsubsection{Conditional Reconnection Probabilities}

The conditional probability that the target user has reconnected with former friend $i$, given he has created a new account \emph{and} that has not been found by stage $t$ when applying search policy $u$, can also be calculated using Bayes' Rule. 
Recall that $A$ is the event that the target user created a new account and $B_i$ is the event that the the target user has reconnected with friend $i$.  Then we have   
\begin{align*}
\P(B_{i}|A,\vu,\vx(t))
& = \frac{\P(x_{i}(t)|B_{i},\vu,A)\P(B_{i}|\vu,A)}%
{\P(x_{i}(t)|B_{i},\vu,A)\P(B_{i}|\vu,A)+(1-\P(B_{i}|\vu,A))}
\\
&=\varphi_{i}\left(
	\frac{
		1-\psi_{i}(\vu,t)
	}
	{
		1-\psi_{i}(\vu,t)\varphi_{i}
	}
\right)
\\
&=
\begin{cases}
\varphi_{i}\left(
\frac{
	N_{i}-x_{i}(t)N_{M}
}{
	N_{i}-\varphi_{i}x_{i}(t)N_{M}
}
\right), & x_{i}(t) = 0,1,\ldots,\left(\left\lceil \frac{N_{i}}{N_{M}} \right\rceil -1 \right)\\
0 & x_{i}(t) = \left\lceil \frac{N_{i}}{N_{M}} \right\rceil 
\end{cases}
\end{align*}

Observe that this probability is the original probability multiplied by the ratio of two linear functions of $x_{i}(t)$.  Because the numerator decreases at a faster rate than the denominator, this probability is strictly decreasing as $x_{i}(t)$ increases from 0 to $\lceil \frac{N_{i}}{N_{M}} \rceil$, provided $\varphi_{i}>0$.  Just as with the conditional existence probability, the monotonicity of this conditional probability matches intuition: the more we query the followers of a certain former friend without finding the target, the less likely it becomes that the target has reconnected with this former friend.

\subsubsection{Distribution of $w(\vx(t),i)$}

The probability of finding the target when querying former friend $i \in \mathcal{V}$ from state $\vx(t)$ is found using the multiplication rule.  Note that the event
\begin{equation*}
\{w(\vx(t),i)=1\} \subseteq B_{i} \subseteq A.
\end{equation*}
Therefore,
\begin{align*}
&\P(w(\vx(t),i)=1) = \P(w(\vx(t),i)=1|B_{i},A,\vx(t))\P(B_{i}|A,\vx(t))\P(A|\vx(t))
\\
& \quad =
\begin{cases}
\varphi_{i}
\left(
	\frac{
		N_{M}
	}{
		N_{i}-\varphi_{i}x_{i}(t)N_{M}
	}
\right)
\left(
	\frac{
		\rho_{0}\prod_{j \in \mathcal{V}}(1-\psi_{j}(\vu,t)\varphi_{j})
	}{
		1-\rho_{0}+\rho_{0}\prod_{j \in \mathcal{V}}(1-\psi_{j}(\vu,t)\varphi_{j})
	}
\right), & 
x_{i}(t)=0,1,\ldots,\left(\left\lceil\frac{N_{i}}{N_{M}}\right\rceil
-2\right)
\\
\varphi_{i}
\left(
\frac{
	N_{i}-x_{i}(t)N_{M}
}{
	N_{i}-\varphi_{i}x_{i}(t)N_{M}
}
\right)
\left(
	\frac{
		\rho_{0}\prod_{j \in \mathcal{V}}(1-\psi_{j}(\vu,t)\varphi_{j})
	}{
		1-\rho_{0}+\rho_{0}\prod_{j \in \mathcal{V}}(1-\psi_{j}(\vu,t)\varphi_{j})
	}
\right), & x_{i}(t) = \left\lceil \frac{N_{i}}{N_{M}} \right\rceil -1.
\end{cases}
\end{align*}

This expression offers several important insights into the dynamics of this search model.  First note that conditioned on the existence of a new target account,
\begin{align}
\P(w(\vx(t),i)=1|A,\vx(t)) &= 
\begin{cases}
\varphi_{i}
\left(
	\frac{
		N_{M}
	}{
		N_{i}-\varphi_{i}x_{i}(t)N_{M}
	}
\right) & x_{i}(t)=0,1,\ldots,\left(\left\lceil\frac{N_{i}}{N_{M}}\right\rceil
-2\right)
\\
\varphi_{i}
\left(
\frac{
	N_{i}-x_{i}(t)N_{M}
}{
	N_{i}-\varphi_{i}x_{i}(t)N_{M}
}
\right)
& 
x_{i}(t)
= \left\lceil\frac{N_{i}}{N_{M}}\right\rceil-1,
\end{cases}
\label{eq: pr_success}
\\[12pt]
\P(w(\vx(t),i)=0|A,\vx(t)) &= 
\begin{cases}
\left(
	\frac{
		N_{i}-\varphi_{i}x_{i}(t+1)N_{M}
	}{
		N_{i}-\varphi_{i}x_{i}(t)N_{M}
	}
\right) & x_{i}(t)=0,1,\ldots,\left(\left\lceil\frac{N_{i}}{N_{M}}\right\rceil
-2\right)
\\
\frac{(1-\varphi_{i})N_{i}}{
	N_{i}-\varphi_{i}x_{i}(t)N_{M}
}
& 
x_{i}(t)
= \left\lceil\frac{N_{i}}{N_{M}}\right\rceil-1. \label{eq: failure}
\end{cases}
\end{align}
We refer to equation \eqref{eq: pr_success} as the probability of success when querying former friend $i$ from state $\vx(t)$.  Likewise, equation \eqref{eq: failure} is the failure probability when querying former friend $i$ from state $\vx(t)$.  Given the target user has created a new account, the success probability for a specific friend $i \in \mathcal{V}$
is strictly increasing as $x_{i}(t)$ increases from 0 to $\left\lceil \frac{N_{i}}{N_{M}}\right\rceil -2$, and is therefore nondecreasing over the corresponding stages $t$.  However, this monotonicity property does not always hold for the final query.  As we have discussed, the final query of $i$ does not necessarily return the same number ($N_{M}$) of results as previous queries of $i$, and has a different functional form for probability of success given in equation \eqref{eq: pr_success}.

\begin{figure}[!hbt]
\centering
\includegraphics[scale = .5]{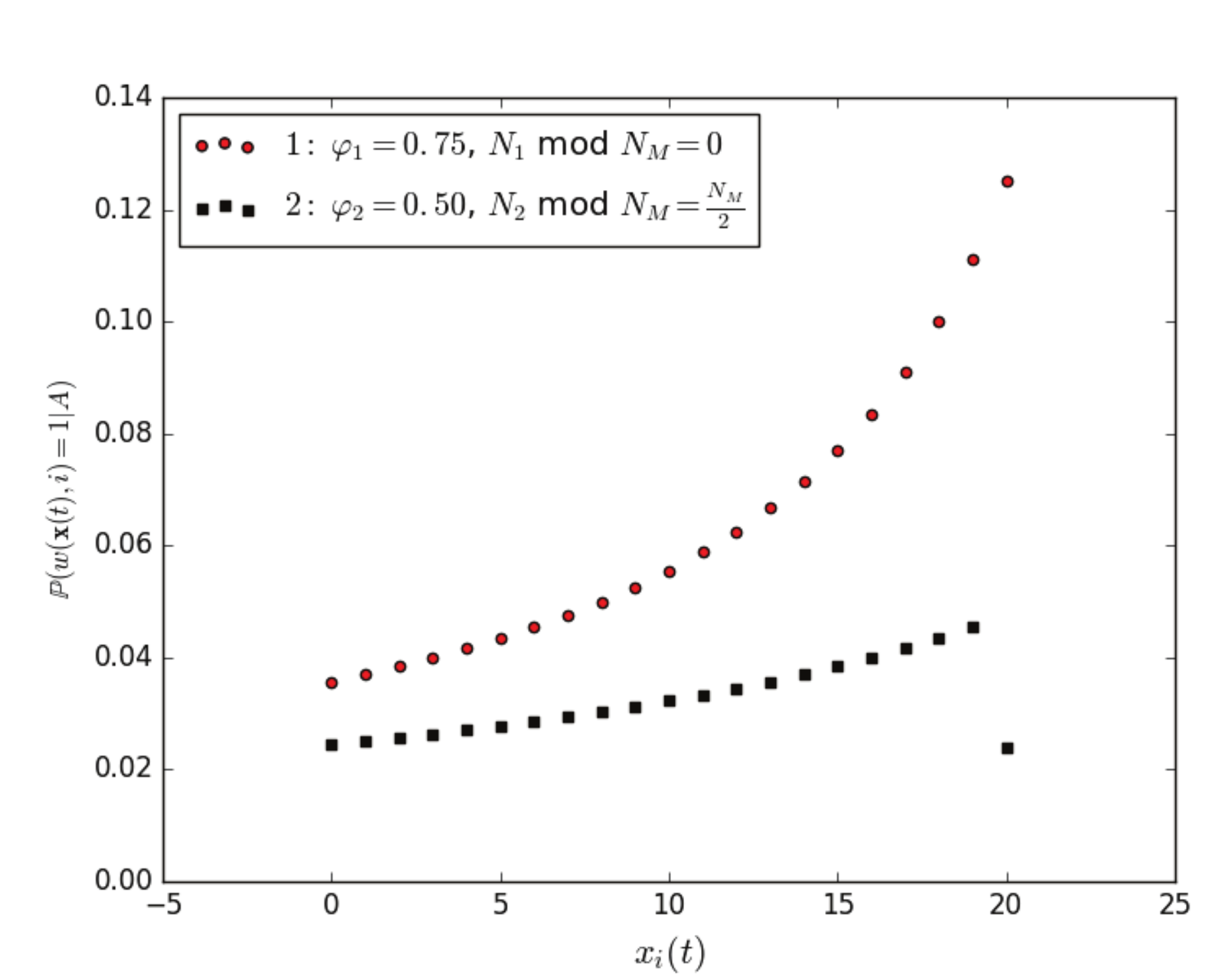}
\caption{Probability of finding the target user's new account, given it exists, as a function of number of queries of former friend $j$.} 
\label{fig: nondecreasing}  
\end{figure}

Figure \ref{fig: nondecreasing} illustrates this monotonicity property for two initial conditions.  In both of the plotted trajectories, $\left\lceil\frac{N_{i}}{N_{M}}\right\rceil=20$.  For former friend 1, $N_{1} \mod N_{M}=0$ and all queries return the same number ($N_{M}$) of results.  In this case the probability of finding the target is strictly increasing over all queries of this former friend's followers.  
The second former friend's success probabilities depicted in Figure \ref{fig: nondecreasing} do not have this characteristic, and the final query returns fewer results than the previous 19 queries.  In this case, we observe that the probability of finding the target is strictly increasing over the first 19 queries, but decreases in the final query because this query returns fewer results.

This monotonicity property is an extension of the monotonicity theorem provided in \cite{urn1}.  As a final note on this property, we observe that this result holds even if we remove the conditioning on $A$.  If in stage $t$ the searcher queried the followers of former friend $i$ and did not find the target user, then in stage $t+1$,
\[
\P(w(\vx(t+1),i)=1)>\P(w(\vx(t),i)=1),
\]
for all $0\leq x_{i}(t+1)<\left\lceil \frac{N_{i}}{N_{M}} \right\rceil-1$, $\varphi_{i}>0$, and $\rho(t)>0$.


\subsection{Analysis: $\bar{\rho}=0$} \label{subsec: analysis}

We provide analysis for the case in which we initialize $\bar{\rho}=0$, i.e., we continue to search until either the target account is found or all follower queries have been exhausted. 
If we were searching for a suspended user's new account, one course of action would be to first execute the query that was most likely to reveal the account.  However, we have shown in \cite{urn1} that this approach does not always yield the optimal policy.  In this section we provide a characterization of the optimal policy that naturally extends from the optimality condition derived in \cite{urn1} for independent urns.

\subsubsection{Expression for Expected Policy Cost}

We now derive an expression for policy cost when $\bar{\rho}=0$.  Let $\vu$ be a valid police and $C_{\vu}$ be the number of unsuccessful queries, or cost of policy $\vu$.  Because $C$ can only take nonnegative integral values $0,1,\ldots,N$,
\begin{align}
\E[C_{\vu}] &= \sum_{t=0}^{N-1}\P(C_{\vu}>t) \nonumber \\
& = \sum_{t=0}^{N-1}\P(C_{\vu}>t|A)\P(A) + 
\sum_{t=0}^{N-1}\P(C_{\vu}>t|A^{c})\P(A^{c})
\nonumber \\
& = \rho_{0}\sum_{t=0}^{N-1}\prod_{k=0}^{t}\P(w(\vx(t),u_{t})=0|A)
+N(1-\rho_{0}). \label{eq: cost0}
\end{align}

The optimal search policy is the valid policy that minimizes this expression.  Formally,
\begin{align}
\vu^{\star}
& = \arg\min_{\vu\in \mathcal{U}}
\E[C_{\vu}] \nonumber
\\
& = 
\arg\min_{\vu\in \mathcal{U}}
\left\{
\rho_{0}\sum_{t=0}^{N-1}\prod_{k=0}^{t}\P(w(\vx(t),u_{t})=0|A)
+N(1-\rho_{0})
\right\} \nonumber
\\
& = 
\arg\min_{\vu\in \mathcal{U}}
\sum_{t=0}^{N-1}\prod_{k=0}^{t}\P(w(\vx(t),u_{t})=0|A)
. \label{eq: cost1}
\end{align}
where $\vu^{\star}$ is the optimal policy and $\mathcal{U}$ is the set of valid policies.  Recall from equation \eqref{eq: u_to_x} that the vectors $\vx(t)$  
can be written as a function of the the search policy.  Not surprisingly, if we commit to exhausting all possible queries in our search for the target, the initial existence probability $\rho_{0}$ does not affect policy optimality.

In order to simplify notation, we define the probability
$
q_{\vu}(t) = 
\P(w(\vx(t),u_{t})=0|A).
$
This is the probability of failing to find the target's new account when executing the $t$th query in policy $\vu$.  This probability is specified in equation \eqref{eq: failure}, and allows us to rewrite the objective function in equation \eqref{eq: cost1} as
\[
\vu^{\star}=\arg\min_{\vu \in \mathcal{U}}\sum_{t=0}^{N-1}\prod_{k=0}^{t}q_{\vu}(t).
\]

\subsubsection{Optimality Conditions}

In \cite{urn1} it is shown that there exists a block policy, in which each urn $i \in \mathcal{V}$ is exhaustively queried before moving on to another urn, that is optimal in any multi-urn search problem.  We now provide an analogous result for this specific application which is proved in Appendix \ref{proof: blockpolicy}.  

\begin{mythm}{\bf (Necessary Conditions for Optimality)} \label{thm: blockpolicy}
If in a suspended user search, follower queries of former friends are executed until either the target user is found or all queries have been exhausted, then any optimal policy must satisfy the following conditions:
\begin{enumerate}
\item[(1)]  The first $\left\lceil \frac{N_{i}}{N_{M}} \right\rceil-1$ queries of each former friend  $i \in \mathcal{V}$ are executed in succession in a single block. 
\item[(2)] For all friends $i \in \mathcal{V}$ such that 
\begin{equation}
\frac{
	N_{i}
}{
	N_{M}\varphi_{i}
}
-\frac{1}{2}\left\lceil
	\frac{N_{i}}{N_{M}}
\right\rceil
>
\frac{
	N_{i}(1-\varphi_{i})
}{
	\varphi_{i}
	\left(
		N_{i}-\left\lceil
			\frac{
				N_{i}
			}{
				N_{M}
			}
		\right\rceil
		N_{M}
		+N_{M}
	\right)
}, \label{eq: condition1}
\end{equation}
all $\left\lceil\frac{N_{i}}{N_{M}}\right\rceil$ queries of $i$'s followers are executed in succession.
\end{enumerate}
\end{mythm}

The first part of Theorem \ref{thm: blockpolicy} follows from the monotonicity of the success probability.  If querying former friend $i$ is optimal in stage $t$, and in the next stage ($t+1$) the success probability for $i$ has increased while success probabilities for all $j \in \mathcal{V}\setminus i$ have remained the same, then intuitively it would be optimal to query $i$ again in stage $t+1$.

The condition in equation \eqref{eq: condition1} is related to how the success probability changes in the final query of each former friend. 
If 
\[
\frac{
	N_{j}
}{
	N_{M}\varphi_{j}
}
-\frac{1}{2}\left\lceil
	\frac{N_{j}}{N_{M}}
\right\rceil
>
\frac{
	N_{j}(1-\varphi_{j})
}{
	\varphi_{j}
	\left(
		N_{j}-\left\lceil
			\frac{
				N_{j}
			}{
				N_{M}
			}
		\right\rceil
		N_{M}
		+N_{M}
	\right)
},
\] then the final query of  $i$ has a lower cost than the previous queries of $i$.  This is the case depicted in Figure \ref{fig: nondecreasing} for former friend $1$.  In this case, querying all of $i$'s followers in succession starting at any stage $t$ is more valuable than executing only the first $\qcount{i}-1$ queries, and any optimal policy will include all of these queries in a single block.   

If on the other hand 
\[
\frac{
	N_{j}
}{
	N_{M}\varphi_{j}
}
-\frac{1}{2}\left\lceil
	\frac{N_{j}}{N_{M}}
\right\rceil
<
\frac{
	N_{j}(1-\varphi_{j})
}{
	\varphi_{j}
	\left(
		N_{j}-\left\lceil
			\frac{
				N_{j}
			}{
				N_{M}
			}
		\right\rceil
		N_{M}
		+N_{M}
	\right)
},
\]
then the final query of former friend $i$ has a higher cost than the previous query.  This is the case of former friend $2$ depicted in Figure \ref{fig: nondecreasing}.  In this case querying all of $i$'s followers in succession starting at any stage $t$ is less beneficial, in terms of minimizing cost, than executing only the first $\qcount{i}-1$ queries.  The optimal policy might separate the final query of $i$ from the first $\qcount{i}-1$ queries in this case.  

If the inequality in equation \eqref{eq: condition1} is instead satisfied with equality, then executing all of the queries of $i$'s followers in succession from any stage $t$ essentially provides the same benefit as executing only the first $\qcount{i}-1$ queries.  In this case, an optimal policy will always exist in which these queries are executed together in a single block, but alternative policies with equal cost might also exist in which the first $\qcount{i}-1$ queries of $i$ are separated from the final query.
%

Theorem \ref{thm: blockpolicy} establishes that the optimal policy is a block policy, but it does not
specify the details of this policy.  The following theorem, which is proved Appendix \ref{proof: optimality}, provides a full characterization of an optimal policy.
\begin{mythm}{\bf (Necessary and Sufficient Conditions for Optimality)} \label{thm: optimality}
In a suspended user search, define 
\begin{align*}
\gamma(\vx(t),i)
&=
\left\{
\begin{array}{l@{\quad}l}
\frac{1}{\varphi_{i}}
\left\lceil
	\frac{N_{i}}{N_{M}}
\right\rceil
-
\frac{
	N_{M}
}{
	2N_{i}
}
\left\lceil
	\frac{N_{i}}{N_{M}}
\right\rceil
\left(
	\left\lceil
		\frac{N_{i}}{N_{M}}
	\right\rceil
	-
	1
\right)
-1,
&
\begin{array}{@{}l}
\frac{
	N_{i}
}{
	N_{M}\varphi_{i}
}
-\frac{1}{2}\left\lceil
	\frac{N_{i}}{N_{M}}
\right\rceil
>
\frac{
	N_{i}(1-\varphi_{i})
}{
	\varphi_{i}
	\left(
		N_{i}-\left\lceil
			\frac{
				N_{i}
			}{
				N_{M}
			}
		\right\rceil
		N_{M}
		+N_{M}
	\right)
}, 
\\
x_{i}(t) = 0,1,\ldots,
\left\lceil
\frac{N_{i}}{N_{M}}
\right\rceil
-1;
\end{array}
\\[36pt]%
\frac{N_{i}}{N_{M}\varphi_{i}}
-\frac{1}{2}\left\lceil
	\frac{N_{i}}{N_{M}}
\right\rceil,
&
\begin{array}{@{}l}
\frac{N_{i}}{N_{M}\varphi_{i}}
-\frac{1}{2}\left\lceil
	\frac{N_{i}}{N_{M}}
\right\rceil
\leq
\frac{
	N_{i}(1-\varphi_{i})
}{
	\varphi_{i}
	\left(
		N_{i}-\left\lceil
			\frac{
				N_{i}
			}{
				N_{M}
			}
		\right\rceil
		N_{M}
		+N_{M}
	\right)
} ,
\\
x_{i}(t) = 0,1,\ldots,
\left\lceil
\frac{N_{i}}{N_{M}}
\right\rceil
-2;
\end{array}
\\[36pt]
\frac{
	N_{i}(1-\varphi_{i})
}{
	\varphi_{i}
	\left(
		N_{i}-\left\lceil
			\frac{
				N_{i}
			}{
				N_{M}
			}
		\right\rceil
		N_{M}
		+N_{M}
	\right)
} ,
&
\begin{array}{@{}l}
\frac{N_{i}}{N_{M}\varphi_{i}}
-\frac{1}{2}\left\lceil
	\frac{N_{i}}{N_{M}}
\right\rceil
\leq
\frac{
	N_{i}(1-\varphi_{i})
}{
	\varphi_{i}
	\left(
		N_{i}-\left\lceil
			\frac{
				N_{i}
			}{
				N_{M}
			}
		\right\rceil
		N_{M}
		+N_{M}
	\right)
} ,
\\
x_{i}(t) = \left\lceil\frac{N_{i}}{N_{M}}\right\rceil -1 ;
\end{array}
\\[36pt]
\infty, & \mathrm{otherwise.}
\end{array}
\right.
\end{align*}
A valid policy is optimal if and only if it satisfies the condition in Theorem \ref{thm: blockpolicy} and it minimizes $\gamma(\vx(t),i)$ in each stage, i.e.,
\[
u_{t}=\arg\min_{i \in \mathcal{V}}\gamma(\vx(t),i) \quad t=0,1,\ldots,N-1.
\]
\end{mythm}
The function $\gamma(\vx(\tau),i)$ arises in the proof of Theorem \ref{thm: optimality} when comparing the costs of  policies which swap the order of querying former friend $i$ with another former friend.  Theorem \ref{thm: optimality} simply says that always choosing the former friend that minimizes $\gamma(\vx(\tau),i)$ produces an optimal policy.  The different cases for $\gamma(\vx(\tau),i)$ correspond to different remaining followers to query of the former friends along with the optimality
conditions from Theorem \ref{thm: blockpolicy}.
 The first case corresponds to the condition in equation \ref{eq: condition1}.  As discussed, this condition implies that executing all queries of $i$ in a single block is more beneficial than executing only the first $\qcount{i}-1$ queries. The other cases follow similar logic: the second case is the value function for the first $\qcount{i}-1$ queries of former friend $i$, and the condition indicates that executing only these queries in a single block is best.  The third case is for the final query of former friend $i$, and the fourth condition sets $\gamma(\vx(t),i)$ to infinity if there are no queries remaining for $i$.

\subsection{Results}

Using the classification results from Section \ref{similarity}, we identified 169 account pairs from our ISIS seed users for testing.  Each pair of accounts consisted of an earlier account, which had been suspended, and an account opened later that belonged to the same user.  Without being able to verify exactly when account suspensions took place, we assumed the later account in each case was opened or used in response to the former account's suspension.  Having collected the friends and followers lists for all of these accounts, we were able to evaluate the performance and effectiveness of the search policy we developed.  
\begin{table}
\centering
\caption{Randomly selected account pairs for testing.} \label{table: test_pairs}
\begin{tabular}{|c|c|c|c|c|c|c|} \hline
Pair & 
Former friends & Reconnection \% & Max Queries $(N)$ \\ \hline
1 & 35 & 40.00\% & 38 \\ \hline 
2 & 310 & 59.68\% & 6609 \\ \hline 
3 & 94 & 17.02\% & 247 \\ \hline 
4 & 87 & 21.84\% & 431 \\ \hline 
5 & 185 & 8.11\% & 198 \\ \hline 
6 & 84 & 22.62\% & 101 \\ \hline 
7 & 63 & 9.52\% & 12007 \\ \hline 
8 & 189 & 4.23\% & 2078 \\ \hline 
9 & 257 & 88.72\% & 4312 \\ \hline 
10 & 109 & 30.28\% & 152 \\ \hline 
11 & 302 & 82.12\% & 5559 \\ \hline 
12 & 344 & 22.67\% & 1314 \\ \hline 
13 & 181 & 9.94\% & 190 \\ \hline 
14 & 87 & 3.45\% & 2965 \\ \hline 
15 & 221 & 2.26\% & 2654 \\ \hline 
\end{tabular}
\end{table}

From the set of 169 account pairs, we randomly chose 15 for testing.  Table \ref{table: test_pairs} shows the number of former friends, the reconnection rate, and the total number of queries possible (or policy length) for each of these account pairs. 
%
%
For each account pair, we identified the friends from the earlier (suspended) account as the ``former friends'' of the subsequent account.  For each of these former friends we determined their reconnection probability using the logistic regression classifier from Section \ref{refollowing}.  We also had the number of followers for each former friend stored in our data set.  We assumed that all of the former friend accounts were still active when the second account was opened.  Finally, we initialized $\rho_{0}=1$.  This initial value is useful because it reduces the expression for expected policy cost to the objective function in equation \eqref{eq: cost1} and allows for direct comparison of actual performance with our theoretical expected number of unsuccessful queries.


In order to evaluate policy performance, we consider the following policies:
\begin{itemize}
	\item Optimal.  This is a policy that minimizes expected cost, found using the necessary and sufficient conditions in Theorem \ref{thm: optimality}.
	\item Greedy.  This policy maximizes the probability of finding the new account at each stage.  Because this probability strictly increases for each former friend $i\in \mathcal{V}$ every time $i$ is queried, excepting the final query of $i$, this policy always meets the necessary condition for optimality given in Theorem \ref{thm: blockpolicy}.  
	\item Min-$N$.  This policy selects the former friend with the minimum number of unqueried followers at each stage.  Because these values strictly decrease for each former friend $i \in \mathcal{V}$ with each query of $i$, this policy always meets the necessary condition for optimality given in Theorem \ref{thm: blockpolicy}.
	\item Max-$P$.  This policy selects the former friend with the highest conditional reconnection probability at each stage.  Because conditional reconnection probabilities strictly decrease for each former friend $i \in \mathcal{V}$ with each query of $i$, this policy does not necessarily meet the conditions in Theorem \ref{thm: optimality}.
	\item Random.  This policy randomly chooses a query from those that are possible at each stage.  
\end{itemize}

\subsubsection{Comparison of Expected Costs}

We computed the expected cost for each policy using equation \eqref{eq: cost0}.  These values do not account for our knowledge of the true reconnections of the second account in each case.  Instead, we assume that our probability model is correct in these computations.

\begin{table}[!hbt]
\centering
\footnotesize
\caption{Cost comparisons for different policies.} \label{table: expected_costs}
\begin{tabular}{c|ccccc||ccccc}
&
\multicolumn{5}{|c||}{Expected Costs} &
\multicolumn{5}{c}{Actual Costs} \\
Pair & Optimal & Greedy & Min-$N$ & Max-$P$ & Random 
&
Optimal & Greedy & Min-$N$ & Max-$P$ & Random \\ \hline
1 & 5.72 & 5.74 & 9.18 & 5.89 & 7.93 & 3.00 & 3.00 & 2.00 & 2.74 & 1.70 \\ 
2 & 2.26 & 2.27 & 4.15 & 88.23 & 68.87 & 0.00 & 0.00 & 2.00 & 44.16 & 20.97 \\ 
3 & 1.22 & 1.22 & 2.00 & 6.09 & 4.91 & 1.00 & 1.00 & 6.00 & 6.28 & 11.63 \\ 
4 & 1.20 & 1.20 & 2.74 & 20.48 & 8.90 & 2.00 & 2.00 & 15.00 & 26.81 & 20.79 \\ 
5 & 2.96 & 2.96 & 9.27 & 3.36 & 6.19 & 5.00 & 5.00 & 15.00 & 6.56 & 10.75 \\ 
6 & 0.96 & 0.96 & 2.52 & 4.43 & 1.86 & 1.00 & 1.00 & 7.00 & 5.53 & 4.49 \\ 
7 & 103.51 & 103.98 & 107.53 & 400.48 & 2170.52 & 5.00 & 5.00 & 12.00 & 283.13 & 1582.28 \\ 
8 & 4.98 & 5.10 & 9.05 & 74.36 & 71.86 & 6.00 & 6.00 & 136.00 & 82.50 & 242.40 \\ 
9 & 2.28 & 2.28 & 4.73 & 80.68 & 57.27 & 0.00 & 0.00 & 2.00 & 54.75 & 5.87 \\ 
10 & 1.01 & 1.01 & 2.99 & 8.64 & 2.27 & 3.00 & 3.00 & 3.00 & 13.76 & 3.27 \\ 
11 & 0.89 & 0.89 & 2.02 & 126.65 & 38.17 & 0.00 & 0.00 & 0.00 & 111.03 & 18.18 \\ 
12 & 2.88 & 2.88 & 6.98 & 42.15 & 18.78 & 0.00 & 0.00 & 6.00 & 44.57 & 19.57 \\ 
13 & 1.50 & 1.50 & 3.82 & 3.26 & 2.52 & 1.00 & 1.00 & 10.00 & 3.41 & 9.57 \\ 
14 & 8.84 & 8.85 & 15.28 & 141.16 & 322.96 & 4.00 & 4.00 & 52.00 & 150.62 & 736.53 \\ 
15 & 1.17 & 1.17 & 2.84 & 61.06 & 20.02 & 7.00 & 7.00 & 61.00 & 143.00 & 390.86 \\ 
\end{tabular}
\end{table}
Table \ref{table: expected_costs} gives the expected costs computed for each policy.  Expected cost values are analytically computed in all cases except for the random policy.  In order to estimate expected cost for a random policy, we generated 500 random policies and computed the expected cost for each.  The average of these 500 expected costs is reported as the random policy expected cost in Table \ref{table: expected_costs}.

The results show that in many cases, the greedy policy and the optimal policy achieve the same cost.  Comparison of these two policies reveals that they are very similar in all cases.  This finding agrees with the findings in \cite{urn1}, which also suggests that there is a bound on the suboptimality of the greedy policy.  The Min-$N$ policy also produces costs close to those of the optimal and greedy policies, while the Max-$P$ and random policies have a substantially higher costs in many cases.

\subsubsection{Comparison of Actual Costs}

In this section we compare the performance of the different policies in finding the target user based on the actual reconnections.  If the target users tended to reconnect in accordance with our probability model we would expect these actual cost values to be similar to the expected costs in Table \ref{table: expected_costs}.  In cases in which the target user reconnected to former friends in a way that would be very unlikely according to our probability model, the actual policy costs might differ substantially from the expected costs.  The actual costs for each policy are reported in Table \ref{table: expected_costs}.  The values reported are the expected number of queries one would have to execute before finding the target user, conditional on the target user's actual  reconnections.  

In some of the 15 cases, the actual costs in Table \ref{table: expected_costs} differ  substantially from the expected costs.  However, the same trend holds: the optimal and greedy policies tend to perform the best, and are nearly indistinguishable in terms of costs.  The Min-$N$ policy performs as well or nearly as well as the optimal policy in some cases, but in a few cases it is much worse.  The Max-$P$ and random policies tend to perform poorly, especially when the target user has not connected with very many former friends (see Table \ref{table: test_pairs}).  Using the optimal or greedy policies can result in substantial cost savings in these cases.  

Account pair 1 provides an example of a case where a random policy can outperform the optimal policy in practice.  The reconnection rate for this target user was 40\% (from Table \ref{table: test_pairs}), but the target did not reconnect with the most probable former friends, according to our probability model (in actuality, it is possible these accounts were suspended when the target opened the new account).  From Table \ref{table: test_pairs}, it is apparent that most of the 35 former friends have fewer than 5,000 followers, because the valid policy length is at most 38 queries.  The random policy performs approximately as we would expect in this case: each random query has approximately a 40\% chance of returning the target.  From the well-known geometric probability distribution, the expected number of failures before the first success is 1.5, which is very close the value reported in Table \ref{table: expected_costs}.

In each of the 15 account pairs, the optimal, greedy and Min-$N$ policies located the target user when querying a former friend that had fewer than 5,000 followers.  For this reason, the actual number of queries in these cases was deterministic, resulting in the integer costs reported in the  Table \ref{table: expected_costs}.  For pair 15, for example, the optimal policy would always find the target user on the 8th query because this is the first former friend in the policy to whom the target user had reconnected, and a single query retrieves all of this former friend's followers.  In this application, many of the former friend accounts have fewer than 5,000 followers and are therefore exhausted in a single followers query.  These accounts, when coupled with a high reconnection probability, are very valuable in a search policy.  Both the greedy and the optimal policies prioritize queries of former friends with relatively high reconnection probabilities and low numbers of followers.

\subsection{Discussion: $\bar{\rho}>0$}

When an existence probability threshold is applied as a termination criterion, Theorems \ref{thm: blockpolicy} and \ref{thm: optimality} no longer hold.  However, the queries that have the highest probability of finding the target user are also the queries that have the largest effect on reducing the conditional existence probability.  We conjecture that the optimal policy in the case for which $\bar{\rho}>0$ will be the same as the optimal policy when $\bar{\rho}=0$ in the initial queries.  At some point, a stage is reached for which a greedy policy becomes more desirable, because it reaches the termination criterion $\rho_{t}<\bar{\rho}$ earlier than a $\bar{\rho}=0$ optimal policy characterized by the conditions in Theorems \ref{thm: blockpolicy} and \ref{thm: optimality}.  


A final consideration for the case in which $\bar{\rho}>0$ involves the initial condition.  The values that conditional existence probability $\rho(t)$ take all depend explicitly on the initial existence probability $\rho_{0}$.  This sensitivity should be explored in analyses or execution of searches that employ this termination criterion.


\section{Conclusion} \label{conclusion}

The growth of online extremism has created the need for capabilities to mitigate the
threat posed by the abusive or threatening behavior of these extremist users.  
In this work we have developed a set of capabilities
which allow for more effective mitigation of these threats.  These capabilities
can be used to enhance the performance of law enforcement or other entities
that are responsible for protecting the public from online extremist groups.
  Our approach combined
statistical modeling of extremist behavior with optimized search policies.
Our behavioral modeling allowed us to predict new extremist users, determine
if two accounts belong to the same extremist user, and predict the network
connections of suspended extremist users when they create new accounts.  We used
our behavioral models to formulate a network search policy to find the new accounts
of suspended extremist users when they return to the social network.  Simulations
based on actual ISIS users found that our policy was much more efficient than
other benchmark approaches.

While our analysis focused on terrorist extremist groups such as ISIS in the social network
Twitter, the capabilities
we developed can apply to any online extremist group and any social network.  Nothing in our modeling
or search policy is specialized to ISIS or Twitter.  Users that engage in some form of online
extremism or harassment will have very similar behavioral characteristics in social networks.  They will
connect to a specific set of users which form their extremist group.  They will
 create new accounts  which will resemble their
old accounts after being suspended.  When they return to the social network after being suspended, they
will reconnect with certain former friends with higher probability.  
In addition, all of our capabilities do not require the cooperation of social network operators.
Therefore, all
the capabilities we developed here are agnostic to the extremist group and social network.

\bibliographystyle{plainnat}
\bibliography{ISIS-exploration}  
\begin{APPENDIX}{} 

\section{Proof of Theorem \ref{thm: blockpolicy}} \label{proof: blockpolicy}
We provide proofs by contradiction that follows the same logic as the block policy proof in \cite{urn1}.  Suppose policy $\vu$ is optimal and does not satisfy condition (1) in Theorem \ref{thm: blockpolicy}.  Then, there exists $i \in \mathcal{V}$ and integers $\tau \geq 0$, $\delta > 1$, and $\Delta > 0$ such that
\begin{align*}
u_{\tau}&=i \\
u_{\tau+\ell} &\neq i \quad \forall \; \ell \in \{1,2,\ldots,\delta-1\} \\
u_{\tau + \delta} &= i \\
u_{\tau+\delta+\Delta} &= i. 
\end{align*}
Note that this final condition simply implies that the query of $i$ in stage $\tau+\delta$ is not the final query of this former friend.  From equation \eqref{eq: failure}, the query failure probabilities in stages $\tau$ and $\tau+\delta$ are
\begin{align*}
q_{\vu}(\tau) 
& = 
\left(
	\frac{
		N_{i}-\varphi_{i}(x_{i}(\tau)+1)N_{M}
	}{
		N_{i}-\varphi_{i}x_{i}(\tau)N_{M}
	}
\right)
\\
q_{\vu}(\tau+\delta) 
& =
\left(
	\frac{
		N_{i}-\varphi_{i}(x_{i}(\tau)+2)N_{M}
	}{
		N_{i}-\varphi_{i}(x_{i}(\tau)+1)N_{M}
	}
\right).
\end{align*}

We construct two alternative policies.  The first alternative policy, $\hat{\vu}$, moves the query of $i$ from stage $\tau+\delta$ to stage $\tau+1$.  The second alternative policy moves the query of $i$ from stage $\tau$ to stage $\tau+\delta-1$.  Each of these alternative policies rearranges the sequence of queries in $\vu$ so that the two queries of former friend $i$ in stages $\tau$ and $\tau+\delta$ are instead executed in succession.  Formally,
\begin{align*}
\hat{u}_{t} &= 
\begin{cases}
	u_{t-1} & t = \tau+1,\ldots,  \tau + \delta\\
	u_{t} & \mathrm{otherwise}
\end{cases}
\\[6pt]
\tilde{u}_{t} & = 
\begin{cases}
	u_{t+1} & t=\tau,\ldots,\tau+\delta-1 \\
	u_{t} & \mathrm{otherwise.}
\end{cases}
\end{align*}

The relationship between the query failure probabilities follows from these policy definitions:
\begin{align*}
q_{\hat{\vu}}(t) &= \begin{cases}
q_{\vu}(\tau+\delta) & t = \tau+1\\
q_{\vu}(t-1) & t=\tau+2,\ldots,\tau+\delta \\
q_{\vu}(t) & \mathrm{otherwise.}
\end{cases}
\\[6pt]
q_{\tilde{\vu}}(t) &= \begin{cases}
q_{\vu}(\tau) & t = \tau+\delta-1\\
q_{\vu}(t+1) & t=\tau,\ldots,\tau+\delta-2 \\
q_{\vu}(t) & \mathrm{otherwise.}
\end{cases}
\end{align*}

We now compare the costs of these policies.  Optimality of $\vu$ implies that the expected cost of policy $\hat{\vu}$ must be at least as high as the cost of $\vu$: 
\begin{align*}
\E[C_{\hat{\vu}}] &\geq \E[C_{\vu}] 
\\%
\sum_{t=0}^{N-1}\prod_{k =0}^{t}q_{\hat{\vu}}(k)
& \geq 
\sum_{t=0}^{N-1}\prod_{k=0}^{t}q_{\vu}(k) 
\\
\sum_{t=\tau+1}^{\tau+\delta}\prod_{k=\tau+1}^{t}q_{\hat{\vu}}(k)
& \geq 
\sum_{t=\tau+1}^{\tau+\delta}\prod_{k=\tau+1}^{t}q_{\vu}(k) 
\\
q_{\hat{\vu}}(\tau+1)
+
q_{\hat{\vu}}(\tau+1)
\sum_{t=\tau+2}^{\tau+\delta}\prod_{k=\tau+2}^{t}q_{\hat{\vu}}(k)
& \geq 
\sum_{t=\tau+1}^{\tau+\delta-1}\prod_{k=\tau+1}^{t}q_{\vu}(k) 
+
 q_{\vu}(\tau+\delta)\prod_{k=\tau+1}^{\tau+\delta-1}q_{\vu}(k)
\\
q_{\vu}(\tau+\delta)
-
 q_{\vu}(\tau+\delta)\prod_{k=\tau+1}^{\tau+\delta-1}q_{\vu}(k)
&\geq
\sum_{t=\tau+1}^{\tau+\delta-1}\prod_{k=\tau+1}^{t}q_{\vu}(k) 
-
q_{\vu}(\tau+\delta)\sum_{t=\tau+1}^{\tau+\delta-1}\prod_{k=\tau+2}^{t}q_{\vu}(k)
\\
\frac{
	q_{\vu}(\tau+\delta)
}{
	1-q_{\vu}(\tau+\delta)
}
&\geq
\frac{
	\sum_{t=\tau+1}^{\tau+\delta-1}\prod_{k=\tau+1}^{t}q_{\vu}(k) 
}{
1-\prod_{k=\tau+1}^{\tau+\delta-1}q_{\vu}(k)
}
\end{align*}

Likewise, optimality of $\vu$ implies that the expected cost of policy $\tilde{\vu}$ must also be at least as high as the cost of $\vu$
\begin{align*}
\E[C_{\tilde{\vu}}] &\geq \E[C_{\vu}] 
\\%
\sum_{t=0}^{N-1}\prod_{k =0}^{t}q_{\tilde{\vu}}(k)
& \geq 
\sum_{t=0}^{N-1}\prod_{k=0}^{t}q_{\vu}(k) 
\\
\sum_{t=\tau}^{\tau+\delta-1}\prod_{k=\tau}^{t}q_{\tilde{\vu}}(k)
& \geq 
\sum_{t=\tau}^{\tau+\delta-1}\prod_{k=\tau}^{t}q_{\vu}(k) 
\\
\sum_{t=\tau}^{\tau+\delta-2}\prod_{k=\tau}^{t}q_{\tilde{\vu}}(k) 
+
 q_{\tilde{\vu}}(\tau+\delta-1)\prod_{k=\tau}^{\tau+\delta-2}q_{\tilde{\vu}}(k)
& 
\geq 
q_{\vu}(\tau)
+
q_{\vu}(\tau)
\sum_{t=\tau+1}^{\tau+\delta-1}\prod_{k=\tau+1}^{t}q_{\vu}(k)
\\
\sum_{t=\tau+1}^{\tau+\delta-1}\prod_{k=\tau+1}^{t}q_{\vu}(k) 
-
q_{\vu}(\tau)\sum_{t=\tau+1}^{\tau+\delta-1}\prod_{k=\tau+2}^{t}q_{\vu}(k)
&\geq
q_{\vu}(\tau)
-
 q_{\vu}(\tau)\prod_{k=\tau+1}^{\tau+\delta-1}q_{\vu}(k)
\\
\frac{
	\sum_{t=\tau+1}^{\tau+\delta-1}\prod_{k=\tau+1}^{t}q_{\vu}(k) 
}{
1-\prod_{k=\tau+1}^{\tau+\delta-1}q_{\vu}(k)
}
&\geq
\frac{
	q_{\vu}(\tau)
}{
	1-q_{\vu}(\tau)
}
\end{align*}

Combining these two conditions, we have
\begin{align*}
\frac{
	q_{\vu}(\tau)
}{
	1-q_{\vu}(\tau)
}
& \leq 
\frac{
	\sum_{t=\tau+1}^{\tau+\delta-1}\prod_{k=\tau+1}^{t}q_{\vu}(k) 
}{
1-\prod_{k=\tau+1}^{\tau+\delta-1}q_{\vu}(k)
}
\leq
\frac{
	q_{\vu}(\tau+\delta)
}{
	1-q_{\vu}(\tau+\delta)
}
\\
\left(
	\frac{
		N_{i}-\varphi_{i}(x_{i}(\tau)+1)N_{M}
	}{
		\varphi_{i}N_{M}
	}
\right)
&
\leq
\frac{
	\sum_{t=\tau+1}^{\tau+\delta-1}\prod_{k=\tau+1}^{t}q_{\vu}(k) 
}{
1-\prod_{k=\tau+1}^{\tau+\delta-1}q_{\vu}(k)
}
\leq
\left(
	\frac{
		N_{i}-\varphi_{i}(x_{i}(\tau)+2)N_{M}
	}{
		\varphi_{i}N_{M}
	}
\right)
\end{align*}
However, under the minimal assumptions that $\varphi_{i}$, $N_{i}$ and $N_{M}$ are positive, the inequality
\[
\left(
	\frac{
		N_{i}-\varphi_{i}(x_{i}(\tau)+1)N_{M}
	}{
		\varphi_{i}N_{M}
	}
\right)
>
\left(
	\frac{
		N_{i}-\varphi_{i}(x_{i}(\tau)+2)N_{M}
	}{
		\varphi_{i}N_{M}
	}
\right)
\]
is strict, which provides a contradiction.

Now suppose optimal policy $\vu$ satisfies condition (1) but does not satisfy condition (2), i.e., there exists a former friend $i \in \mathcal{V}$ for which
\[
\frac{
	N_{i}
}{
	N_{M}\varphi_{i}
}
-\frac{1}{2}\left\lceil
	\frac{N_{i}}{N_{M}}
\right\rceil
>
\frac{
	N_{i}(1-\varphi_{i})
}{
	\varphi_{i}
	\left(
		N_{i}-\left\lceil
			\frac{
				N_{i}
			}{
				N_{M}
			}
		\right\rceil
		N_{M}
		+N_{M}
	\right)
}
\]
that is not queried in a single block.  Let $\tau-\qcount{i}+2$ be the first stage in policy $\vu$ in which former friend $i$ is queried.  Because the policy satisfies condition (1), it follows that 
\[
u_{t}=i \quad t=\tau-\qcount{i}+2,\tau-\qcount{i}+1,\ldots,\tau.
\]
Also, let $\tau+\delta$ be the stage corresponding to the final query of former friend $i$.  By assumption this final query is not executed in succession with the first $\qcount{i}-1$ queries of $i$, so $\delta>1$.

As in the previous part of the proof, we define two alternative policies, each moving two final queries of $i$ into a single block.  
\begin{align*}
\hat{u}_{t} &= 
\begin{cases}
	u_{t-1} & t = \tau+1,\ldots, \tau + \delta\\
	u_{t} & \mathrm{otherwise}
\end{cases}
\\[6pt]
\tilde{u}_{t} & = 
\begin{cases}
	u_{t+\qcount{i}-1} & t=\tau-\qcount{i}+2,\ldots,\tau-\qcount{i}+\delta \\
	i & t= \tau-\qcount{i}+\delta+1,\ldots,\tau+\delta\\
	u_{t} & \mathrm{otherwise.}
\end{cases}
\end{align*}

The relationship between the query failure probabilities follows from these policy definitions:
\begin{align*}
q_{\hat{\vu}}(t) &= \begin{cases}
q_{\vu}(\tau+\delta) & t = \tau+1\\
q_{\vu}(t-1) & t=\tau+2,\ldots,\tau+\delta \\
q_{\vu}(t) & \mathrm{otherwise.}
\end{cases}
\\[6pt]
q_{\tilde{\vu}}(t) &= \begin{cases}
q_{\vu}(t+\qcount{i}-1) & t = \tau-\qcount{i}+2,\ldots,\tau-\qcount{i}+\delta\\
q_{\vu}(t-\delta+1) & t=\tau-\qcount{i}+\delta+1,\ldots,\tau+\delta-1 \\
q_{\vu}(t) & \mathrm{otherwise.}
\end{cases}
\end{align*}

Note also that, from equation \eqref{eq: failure},
\begin{align*}
\prod_{k=\tau-\qcount{i}+2}^{\tau-\qcount{i}+2+t} q_{\vu}(k) 
& = 
\frac{N_{i}-\varphi_{i}(t+1)N_{M}}{N_{i}},
\quad t=0,1,\ldots,\qcount{i}-2 
\\
\prod_{t=\tau-\qcount{i}+2}^{\tau}q_{\vu}(t) 
& = 
\frac{
	N_{i}-\varphi_{i}\left(\qcount{i}-1\right)N_{M}
}{
	N_{i}
}
\\
\sum_{t=\tau-\qcount{i}+2}^{\tau}
\
\prod_{k=\tau-\qcount{i}+2}^{t} q_{\vu}(k)
&=
\left(
	\qcount{i}-1
\right)
-
\frac{\varphi_{i}N_{M}}{2N_{i}}
\left(
	\qcount{i}
\right)
\left(
	\qcount{i}-1
\right)
\\
q_{\vu}(\tau+\delta)
& = 
\frac{(1-\varphi_{i})N_{i}}{
	N_{i}-\varphi_{i}\left(\qcount{i}-1\right)N_{M}
}.
\end{align*}

As in the previous part of the proof, we compare the costs of the policies.  
\begin{align*}
\E[C_{\hat{\vu}}] &\geq \E[C_{\vu}] 
\\%
\sum_{t=0}^{N-1}\prod_{k =0}^{t}q_{\hat{\vu}}(k)
& \geq 
\sum_{t=0}^{N-1}\prod_{k=0}^{t}q_{\vu}(k) 
\\
\sum_{t=\tau+1}^{\tau+\delta}\prod_{k=\tau+1}^{t}q_{\hat{\vu}}(k)
& \geq 
\sum_{t=\tau+1}^{\tau+\delta}\prod_{k=\tau+1}^{t}q_{\vu}(k) 
\\
q_{\hat{\vu}}(\tau+1)
+
q_{\hat{\vu}}(\tau+1)
\sum_{t=\tau+2}^{\tau+\delta}\prod_{k=\tau+2}^{t}q_{\hat{\vu}}(k)
& \geq 
\sum_{t=\tau+1}^{\tau+\delta-1}\prod_{k=\tau+1}^{t}q_{\vu}(k) 
+
 q_{\vu}(\tau+\delta)\prod_{k=\tau+1}^{\tau+\delta-1}q_{\vu}(k)
\\
q_{\vu}(\tau+\delta)
-
 q_{\vu}(\tau+\delta)\prod_{k=\tau+1}^{\tau+\delta-1}q_{\vu}(k)
&\geq
\sum_{t=\tau+1}^{\tau+\delta-1}\prod_{k=\tau+1}^{t}q_{\vu}(k) 
-
q_{\vu}(\tau+\delta)\sum_{t=\tau+1}^{\tau+\delta-1}\prod_{k=\tau+2}^{t}q_{\vu}(k)
\\
\frac{
	q_{\vu}(\tau+\delta)
}{
	1-q_{\vu}(\tau+\delta)
}
&\geq
\frac{
	\sum_{t=\tau+1}^{\tau+\delta-1}\prod_{k=\tau+1}^{t}q_{\vu}(k) 
}{
1-\prod_{k=\tau+1}^{\tau+\delta-1}q_{\vu}(k)
}
\end{align*}

Likewise, optimality of $\vu$ implies that the expected cost of policy $\tilde{\vu}$ must also be at least as high as the cost of $\vu$:
\begin{align*}
&\E[C_{\tilde{\vu}}] \geq \E[C_{\vu}] 
\\%
&\sum_{t=0}^{N-1}\prod_{k =0}^{t}q_{\tilde{\vu}}(k)
\geq 
\sum_{t=0}^{N-1}\prod_{k=0}^{t}q_{\vu}(k) 
\\[6pt]
&\sum_{t=\tau-\qcount{i}+2}^{\tau+\delta-1}
\
\prod_{k=\tau-\qcount{i}+2}^{t}q_{\tilde{\vu}}(k)
 \geq 
\sum_{t=\tau-\qcount{i}+2}^{\tau+\delta-1}
\
\prod_{k=\tau-\qcount{i}+2}^{t}q_{\vu}(k) 
\\[6pt]
&\sum_{t=\tau-\qcount{i}+2}^{\tau-\qcount{i}+\delta}
\
\prod_{k=\tau-\qcount{i}+2}^{t}q_{\tilde{\vu}}(k) 
+
 \prod_{k=\tau-\qcount{i}+2}^{\tau-\qcount{i}+\delta}q_{\tilde{\vu}}(k)
 \sum_{t=\tau-\qcount{i}+\delta+1}^{\tau+\delta-1}
 \
 \prod_{k=\tau-\qcount{i}+\delta+1}^{t}q_{\tilde{\vu}}(k)
\\  
& \qquad \geq 
\sum_{t=\tau-\qcount{i}+2}^{\tau}
\
\prod_{k=\tau-\qcount{i}+2}^{t}q_{\vu}(k) 
+
 \prod_{k=\tau-\qcount{i}+2}^{\tau}q_{\vu}(k)
 \sum_{t=\tau+1}^{\tau+\delta-1}\prod_{k=\tau+1}^{t}q_{\vu}(k)
\\[6pt]
&\sum_{t=\tau+1}^{\tau+\delta-1}\prod_{k=\tau+1}^{t}q_{\vu}(k) 
+
 \prod_{k=\tau+1}^{\tau+\delta-1}q_{\vu}(k)
 \sum_{t=\tau-\qcount{i}+2}^{\tau}
 \
 \prod_{k=\tau-\qcount{i}+2}^{t}q_{\vu}(k)
\\  
& \qquad \geq 
\sum_{t=\tau-\qcount{i}+2}^{\tau}
\
\prod_{k=\tau-\qcount{i}+2}^{t}q_{\vu}(k) 
+
 \prod_{k=\tau-\qcount{i}+2}^{\tau}q_{\vu}(k)
 \sum_{t=\tau+1}^{\tau+\delta-1}\prod_{k=\tau+1}^{t}q_{\vu}(k)
\\[6pt]
&\frac{
	\sum_{t=\tau+1}^{\tau+\delta-1}\prod_{k=\tau+1}^{t}q_{\vu}(k)  
}{
1-\prod_{k=\tau+1}^{\tau+\delta-1}q_{\vu}(k)
}
\geq
\frac{
	\sum_{t=\tau-\qcount{i}+2}^{\tau}\prod_{k=\tau-\qcount{i}+2}^{t}q_{\vu}(k) 
}{
	1-\prod_{k=\tau-\qcount{i}+2}^{\tau}q_{\vu}(k)
}.
\end{align*}

Combining these two conditions, we have
\[
\frac{
	\sum_{t=\tau-\qcount{i}+2}^{\tau}\prod_{k=\tau-\qcount{i}+2}^{t}q_{\vu}(k) 
}{
	1-\prod_{k=\tau-\qcount{i}+2}^{\tau}q_{\vu}(k).
}
 \leq 
\frac{
	\sum_{t=\tau+1}^{\tau+\delta-1}\prod_{k=\tau+1}^{t}q_{\vu}(k) 
}{
1-\prod_{k=\tau+1}^{\tau+\delta-1}q_{\vu}(k)
}
\leq
\frac{
	q_{\vu}(\tau+\delta)
}{
	1-q_{\vu}(\tau+\delta)
}.
\]
This inequality implies that
\begin{align*}
\frac{
	\sum_{t=\tau-\qcount{i}+2}^{\tau}\prod_{k=\tau-\qcount{i}+2}^{t}q_{\vu}(k) 
}{
	1-\prod_{k=\tau-\qcount{i}+2}^{\tau}q_{\vu}(k).
}
&
\leq
\frac{
	q_{\vu}(\tau+\delta)
}{
	1-q_{\vu}(\tau+\delta)
}
\\
\frac{N_{i}}{\varphi_{i}N_{M}}-\frac{1}{2}\qcount{i}
& 
\leq
\frac{
	(1-\varphi_{i})N_{i}
}{
	\varphi_{i}\left(N_{i}-\qcount{i}N_{M}+N_{M}\right),
}
\end{align*}
which is a contradiction.


\section{Proof of Theorem \ref{thm: optimality}}\label{proof: optimality}
First observe that a policy satisfying the condition in Theorem \ref{thm: optimality} always exists.  Such a policy can be constructed algorithmically by picking the former friend $i: i = \arg\min_{j\in\mathcal{V}}\gamma(\vx(t),j)$ and querying $i$ successively until a stage $t'$ is reached for which $\gamma(\vx(t'),i)\neq\gamma(\vx(t),i)$.  At this stage a new former friend is chosen for querying according to the same criterion.   

An important characteristic of $\gamma(\vx(t),j)$  is that it is nondecreasing in $t$ for all $j \in \mathcal{V}$.    
This property implies that for any policy $\vu$ that satisfies the condition in Theorem \ref{thm: optimality}, $\gamma(\vx(t),u_{t})\leq \gamma(\vx(t+1),u_{t+1})$.

We now show by contradiction that a policy which does not meet the condition of Theorem \ref{thm: optimality} cannot be optimal.  We consider only policies that meet the condition of Theorem \ref{thm: blockpolicy}, as we have shown this condition to be necessary for optimality.  Suppose optimal policy $\vu$ meets the necessary condition for optimality in Theorem \ref{thm: blockpolicy} but does not meet the condition of Theorem \ref{thm: optimality}.  Then, there must be at least one stage $\tau$ in which $\gamma(\vx(\tau),u_{\tau}) > \gamma(\vx(\tau+1),u_{\tau+1})$.  Because $\gamma(\vx(t),j)$ is nondecreasing in $t$ for all $j$, this condition implies $u_{\tau} \neq u_{\tau+1}$.  For clarity of notation, assume that $u_{\tau}=i$ and $u_{\tau+1}=j$.

We construct an alternate policy in which the order of these former friends $i$ and $j$ is reversed.  Let $\ell$ be the earliest stage for which $\gamma(\vx(\ell),i)=\gamma(\vx(\tau),i)$ and
\[
u_{t}=i \quad \forall t \in \{\ell,\ell+1,\ldots,\tau\}.
\]
Also let $L$ be the latest stage for which $\gamma(\vx(L),j)=\gamma(\vx(\tau+1),j)$ and
\[
u_{t}=j \quad \forall t \in \{\tau+1,\tau+2,\ldots,L\}.
\]
Let $\delta=\tau-\ell+1$ be the number of successive stages that $i$ is queried in this sequence and let $\Delta=L-\tau$ be the number of successive stages that $j$ is queried in this sequence.  In our alternative policy $\tilde{\vu}$, we let
\[
\tilde{u}_{t}=\begin{cases}
u_{t} & t=0,\ldots,\ell-1,L+1,\ldots,N-1 \\
j & t \in \ell,\ell+1,\ldots,\ell+\Delta-1 \\
i & t \in \ell+\Delta,\ell+\Delta+2,\ldots,L.
\end{cases}
\]  
This relationship implies
\[
q_{\tilde{\vu}}(t)=\begin{cases}
q_{\vu}(t) & t=0,\ldots,\ell-1,L+1,\ldots,N-1 \\
q_{\vu}(t+\delta) & t \in \ell,\ell+1,\ldots,\ell+\Delta-1 \\
q_{\vu}(t-\Delta) & t \in \ell+\Delta,\ell+\Delta+2,\ldots,L.
\end{cases}
\]

From the optimality of $\vu$, we have
\begin{align}
\E[C_{\tilde{\vu}}]&\geq \E[C_{\vu}] \nonumber \\
\sum_{t=0}^{N-1}\prod_{k=0}^{t}q_{\tilde{\vu}}(k) 
&\geq
\sum_{t=0}^{N-1}\prod_{k=0}^{t}q_{\vu}(k) 
\nonumber \\%
\sum_{t=\ell}^{L}\prod_{k=\ell}^{t}q_{\tilde{\vu}}(k) 
&\geq
\sum_{t=\ell}^{L}\prod_{k=\ell}^{t}q_{\vu}(k) 
\nonumber \\%
\sum_{t=\ell}^{\ell+\Delta-1}\prod_{k=\ell}^{t}q_{\tilde{\vu}}(k) 
+
\left(\prod_{k=\ell}^{\ell+\Delta-1}q_{\tilde{\vu}}(k)\right)\sum_{t=\ell+\Delta}^{L}\prod_{k=\ell+\Delta}^{t}q_{\tilde{\vu}}(k) 
&\geq
\sum_{t=\ell}^{\tau}\prod_{k=\ell}^{t}q_{\vu}(k) 
+
\left(\prod_{k=\ell}^{\tau}q_{\vu}(k)\right)\sum_{\tau+1}^{L}\prod_{k=\tau+1}^{t}q_{\vu}(k) 
\nonumber \\%
\sum_{t=\tau+1}^{L}\prod_{k=\tau+1}^{t}q_{\vu}(k) 
+
\left(\prod_{k=\tau+1}^{L}q_{\vu}(k)\right)\sum_{t=\ell}^{\tau}\prod_{k=\ell}^{t}q_{\vu}(k) 
&\geq
\sum_{t=\ell}^{\tau}\prod_{k=\ell}^{t}q_{\vu}(k) 
+
\left(\prod_{k=\ell}^{\tau}q_{\vu}(k)\right)\sum_{\tau+1}^{L}\prod_{k=\tau+1}^{t}q_{\vu}(k) 
\nonumber \\%
\frac{
	\sum_{t=\tau+1}^{L}\prod_{k=\tau+1}^{t}q_{\vu}(k) 
}
{
	1-\prod_{k=\tau+1}^{L}q_{\vu}(k)
}
&\geq
\frac{
	\sum_{t=\ell}^{\tau}\prod_{k=\ell}^{t}q_{\vu}(k) 
}{
	1-\prod_{k=\ell}^{\tau}q_{\vu}(k)
}. \label{eq: comparison}
\end{align}

We have multiple cases to consider when comparing these policy costs.  Consider the sequence of queries of former friend $i$, starting in  stage $\ell$ and ending in stage $\tau$.  Our assumptions on policy $\vu$ (that it satisfies Theorem \ref{thm: blockpolicy}) and our method of selecting stage $\ell$ allow for three distinct possibilities:

\begin{list}{\bf Case \arabic{enumi}.}{\usecounter{enumi}}

\item Stage $\ell$ is the first query of $i$ in policy $\vu$ and stage $\tau$ is the $\left(\qcount{i}-1\right)$th query of $i$.  By adhering to the necessary conditions for optimality in Theorem \ref{thm: blockpolicy}, this case implies that 
$
\frac{N_{i}}{\varphi_{i}N_{M}}-\frac{1}{2}\qcount{i}
\leq
\frac{
	(1-\varphi_{i})N_{i}
}{
	\varphi_{i}\left(N_{i}-\qcount{i}N_{M}+N_{M}\right).
}
$
Observe that in this case the quantity
\begin{align*}
\frac{
	\sum_{t=\ell}^{\tau}\prod_{k=\ell}^{t}q_{\vu}(k) 
}{
	1-\prod_{k=\ell}^{\tau}q_{\vu}(k)
}
& =
\frac{
	\left(\qcount{i}-1\right)-\frac{\varphi_{i}N_{M}}{2N_{i}}\left(\qcount{i}-1\right)\qcount{i}
}{
	\frac{\varphi_{i}\left(\qcount{i}-1\right)N_{M}}{N_{i}}
}\\
& = 
\frac{N_{i}}{\varphi_{i}N_{M}}-\frac{1}{2}\qcount{i}
\\
& = \gamma(\vx(\tau),i)
\end{align*}

\item Stage $\ell<\tau$ is the first query of $i$ in policy $\vu$ and stage $\tau$ is the final (or $\qcount{i}$th) query of $i$.  Equality of $\gamma(\vx(t),i)$  across these stages implies
$
\frac{N_{i}}{\varphi_{i}N_{M}}-\frac{1}{2}\qcount{i}
\geq
\frac{
	(1-\varphi_{i})N_{i}
}{
	\varphi_{i}\left(N_{i}-\qcount{i}N_{M}+N_{M}\right).
}
$
Observe that in this case the quantity
\begin{align*}
\frac{
	\sum_{t=\ell}^{\tau}\prod_{k=\ell}^{t}q_{\vu}(k) 
}{
	1-\prod_{k=\ell}^{\tau}q_{\vu}(k)
}
& =
\frac{
	\left(\qcount{i}-1\right)-\frac{\varphi_{i}N_{M}}{2N_{i}}\left(\qcount{i}-1\right)\qcount{i} + (1-\varphi_{i})
}{
	\varphi_{i}
}\\
& = 
\frac{1}{\varphi_{j}}
\left\lceil
	\frac{N_{j}}{N_{M}}
\right\rceil
-
\frac{
	N_{M}
}{
	2N_{j}
}
\left\lceil
	\frac{N_{j}}{N_{M}}
\right\rceil
\left(
	\left\lceil
		\frac{N_{j}}{N_{M}}
	\right\rceil
	-
	1
\right)
-1
\\
& = \gamma(\vx(\tau),i)
\end{align*}

\item Stage $\ell=\tau$ is the final query of $i$ in policy $\vu$.  If this is the case, then $\gamma(\vx(\tau),i)=\frac{N_{i}(1-\varphi_{i})}{\varphi_{i}\left(N_{i}-\qcount{i}N_{M}+N_{M}\right)}$, irrespective of whether this query is the only query of $i$.  
Observe that in this final case,
\begin{align*}
\frac{
	\sum_{t=\ell}^{\tau}\prod_{k=\ell}^{t}q_{\vu}(k) 
}{
	1-\prod_{k=\ell}^{\tau}q_{\vu}(k)
}
& =
\frac{\left(
	\frac{(1-\varphi_{i})N_{i}}{
		N_{i}-\varphi_{i}\left(\qcount{i}-1\right)N_{M}
	}
\right)}{\left(
	\frac{
		N_{i}-\varphi_{i}\left(\qcount{i}-1\right)N_{M}-(1-\varphi_{i})N_{i}
	}
	{
		N_{i}-\varphi_{i}\left(\qcount{i}-1\right)N_{M}
	}
\right)}\\
& = 
\frac{
	(1-\varphi_{i})N_{i}
}{
	\varphi_{i}\left(N_{i}- \qcount{i}+N_{M}\right)
}
\\
& = \gamma(\vx(\tau),i).
\end{align*}
Note that if stage $\tau$ corresponds to the \emph{only} query of $i$ in policy $\vu$, then $\qcount{i}=1$ and this expression reduces to $\frac{1-\varphi_{i}}{\varphi_{i}}$, which is correct.
\end{list}

These three cases similarly apply to the sequence of queries of former friend $j$ in stages $\tau+1,\ldots,L$.  Therefore, in all cases equation \eqref{eq: comparison} reduces to
\begin{align*}
\frac{
	\sum_{t=\tau+1}^{L}\prod_{k=\tau+1}^{t}q_{\vu}(k) 
}
{
	1-\prod_{k=\tau+1}^{L}q_{\vu}(k)
}
&\geq
\frac{
	\sum_{t=\ell}^{\tau}\prod_{k=\ell}^{t}q_{\vu}(k) 
}{
	1-\prod_{k=\ell}^{\tau}q_{\vu}(k)
}
\\
\gamma(\vx(\tau+1),j) 
& \geq 
\gamma(\vx(\tau),i),
\end{align*}
which is a contradiction and shows that Theorem \ref{thm: optimality} provides a necessary condition for optimality.  

To show that this condition is sufficient for optimality, suppose now that policy $\vu$ satisfies the conditions in Theorems \ref{thm: blockpolicy} and \ref{thm: optimality}, but that it is not optimal.  This assumption implies that there is another policy, $\vu^{\star}$ with a lower cost.  From our previous arguments, $\vu^{\star}$ must also satisfy the conditions in Theorems \ref{thm: blockpolicy} and \ref{thm: optimality}.  Because $\gamma(\vx(t),i)$ is nondecreasing in $t$ for all $i \in \mathcal{V}$, the possible differences between the policies $\vu$ and $\vu^{\star}$ are in stages where ties exist, i.e., 
\[
\exists \; i,j \in \mathcal{V}: i \neq j, \; \gamma(\vx(t),i)=\gamma(\vx(t),j).
\]
However, it follows from our development above that these reorderings do not result in a change in cost.  In other words, all policies that meet the conditions in Theorems \ref{thm: blockpolicy} and \ref{thm: optimality} result in the same cost and therefore must be optimal.

\section{Classification Threshold Sensitivity}\label{app:sensitivity}

We provide a brief discussion of the sensitivity of the results to changes in the classification threshold.  In the previous section, we selected threshold $P=0.782$ based on the shape of the ROC curve and our desire to keep the number of false positive classifications low.  We now consider how different values of threshold $P$ affect the ``paired accounts'' graph  depicted in Figure \ref{merge-graph}.

\begin{figure}[!hbt]
\centering
\includegraphics[scale = 0.5]{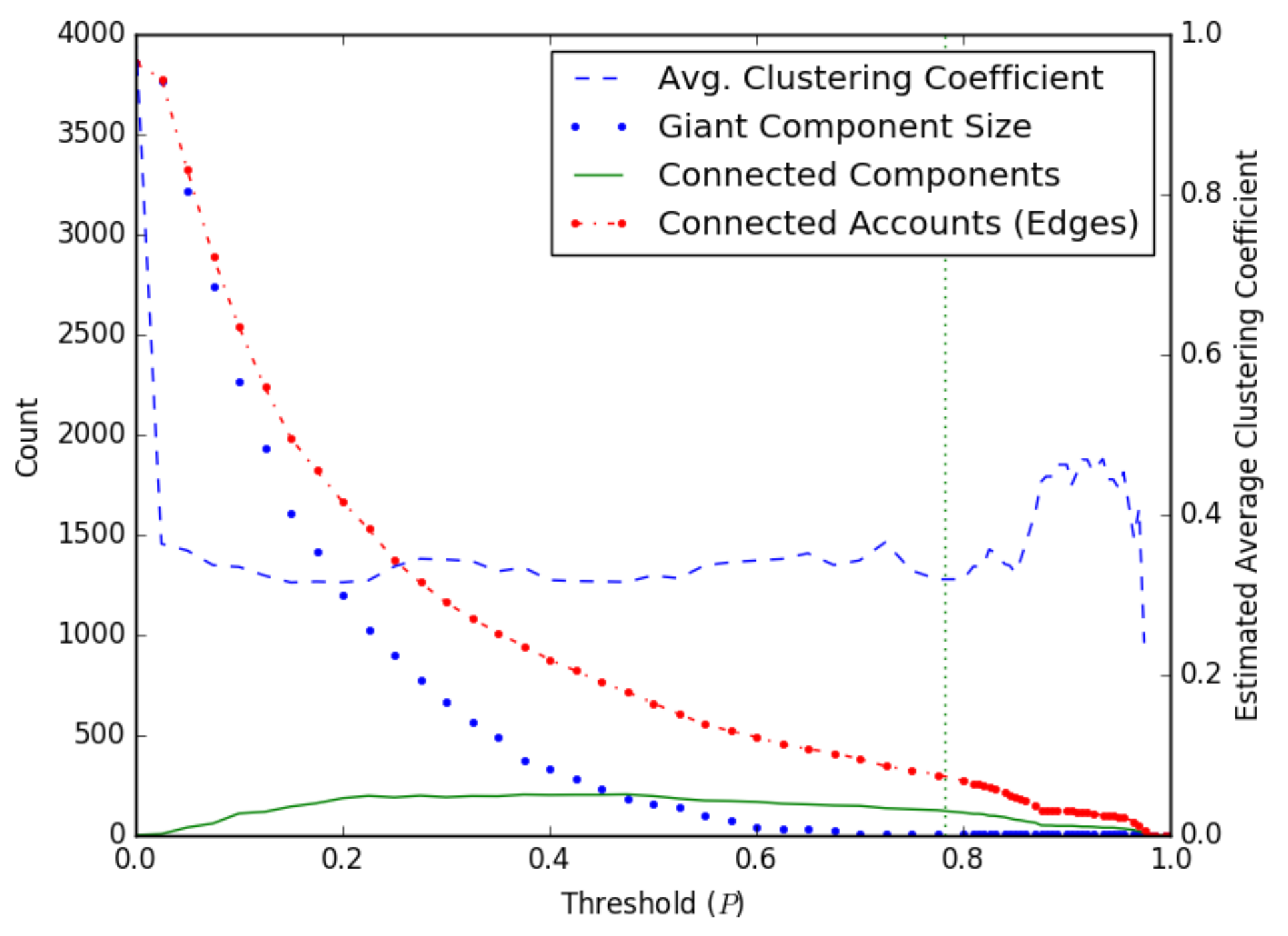}
\caption{Paired accounts graph properties as a function of threshold $P$.  The threshold value 0.782 from
equation \eqref{classifier} is indicated on the plot.}\label{sensitivity}
\end{figure}

Figure \ref{sensitivity} gives several properties of the ``paired account'' graph as a function of $P$.  As we would expect, when our classification threshold $P=0$ the graph is fully connected, which indicates that all accounts are classified as belonging to the same user.  As $P$ increases, the number of connected accounts and the size of the giant component decrease rapidly.  Of interest is the estimated average clustering coefficient, measured on the right-hand scale in Figure \ref{sensitivity}.  If we had access to the true classifications so that we could produce a graph of connected accounts that belonged to the same users, each component would be fully connected.  Average clustering provides a measure of how much a graph exhibits this property by estimating how often a triad of connected nodes is fully connected.  

We see from Figure \ref{sensitivity}  that the average clustering coefficient is relatively stable for a wide range of threshold values, but as $P$ increases beyond approximately 0.85 we observe an increase in the average clustering coefficient that suggests that there are clusters of profiles in our data that are all very similar.  Component A, indicated in Figure \ref{merge-graph} and enumerated in Table \ref{comp-A}, is an example of such a cluster.  There are other fully connected clusters in Figure \ref{merge-graph} consisting of more nodes.  These clusters represent users who open many Twitter accounts and retain very similar profile features.  Further investigation of these accounts reveals that they are nearly all suspended, suggesting that account suspensions are the driving force behind the creation of these multiple accounts.  As noted earlier, in at least some cases these accounts are created by high-profile jihadists.

Decreasing $P$ from 0.782 appears to rapidly increase the number of false positive classifications.  This result becomes quickly apparent in the appearance of a large but loosely connected component in the paired graph structure.  For example, reducing the classification threshold to $P=0.668$ (indicated on the ROC plot in Figure \ref{L1ROC}) increases the profile pairs classified as belonging to the same user to 455.  In many cases, these additional pairs appear to be correct classifications.  For example, component B in Figure \ref{merge-graph} appears as a fully connected component using this threshold.  However, we also observe the formation of the loosely connected component indicated as ``component C'' in Figure \ref{merge-low}.  Table \ref{comp-C} shows the profile features for the accounts comprising this component, which appear to belong to several different users.

\begin{figure}
\centering
\fbox{
\includegraphics[scale=0.25]{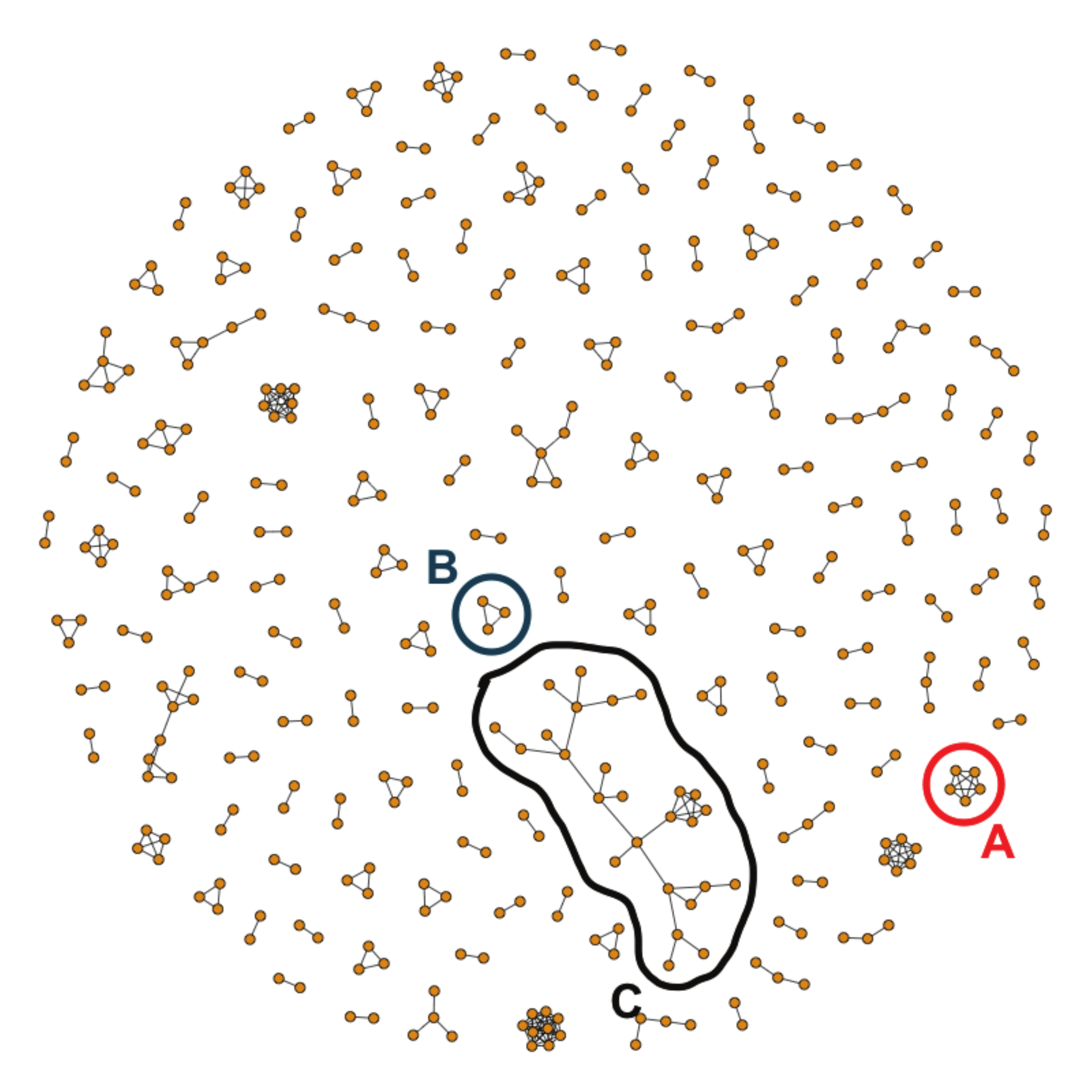}
}
\caption{Graph representation of accounts belonging to the same user using our regression model and equation \eqref{classifier} with a threshold of 0.668.} \label{merge-low}
\end{figure}

\begin{table} \centering
\caption{Accounts comprising component C.} \label{comp-C}
\begin{tabular}{|l|l|l|} 
\multicolumn{3}{c}{} \\ \hline
Screen Name & Name & Profile Pic \\ \hline
AAbuAAwlaki & Abu Awlaki & [None]\\ 
abu\_alia2 & abu alia & [None]\\ 
Abdullah4510394 & Abdullah  & [None]\\ 
abu\_abdillah12 & Abu Abdullah & [None]\\ 
dewdropz69 & Abdullah & [None]\\ 
Ummabdullaa & Umm Abdullah & [None]\\ 
abouabdullah7 & abou abdullah  & ff\ldots{}e0\\ 
AbuAbdullah1400 & Abu Abdullah & ff\ldots{}ff\\ 
abouosama6 & Abouosama & [None]\\ 
Abuusamah17 & Abu usamah & [None]\\ 
AbuIabulfida & Abu Abdullah & e1\ldots{}00\\ 
AbuAyman2011 & Abu Ayman & [None]\\ 
AbuMuhammad1503 & Abu Muhammad & [None]\\ 
abu\_malhama4 & Abu Malhama & [None]\\ 
moabibkhab & abu hamad & [None]\\ 
nahida\_muhammad & Nahida muhammad & [None]\\ 
abumusab\_musab & Abu musab & [None]\\ 
xcon\_cp\_dc & Abu Musa & [None]\\ 
AbuSaalihah06 & Abu Saalihah  & [None]\\ 
AbuSaalihah07 & Abu Saalihah  & 00\ldots{}00\\ 
AbuSaalihah08 & Abu Saalihah & 00\ldots{}00\\ 
AbuSaalihah13 & Abu Saalihah  & 00\ldots{}00\\ 
Abu\_swaaliha & abu swaaliha  & 1e\ldots{}c3\\ 
Abu\_Malhama5 & Abu Malhama & bf\ldots{}00\\ 
omertalhaa & Abu Talha & [None]\\ 
islamobjective & Abu Ramadi & [None]\\ \hline
\end{tabular}
\end{table}

\section{Features for Refollowing Model}\label{app:features_list}
The complete list of features used in the refollow model from Section \ref{refollowing}
is listed below.

\begin{itemize}
\item {\bf Friend}'s number of Twitter friends (Log).
\item {\bf Friend}'s number of Twitter followers (Log).
\item {\bf Friend}'s number of Tweets (Log).
\item Account age difference between {\bf Friend} and {\bf User0}.
\item Binary indicator of whether {\bf Friend} was following {\bf User0}.
\item Number of times {\bf User0} mentioned {\bf Friend} in a tweet (Log).
\item Number of times {\bf User0} retweeted one of {\bf Friend}'s tweets (Log).
\item Number of times {\bf User0} replied to one of {\bf Friend}'s tweets (Log).
\item {\bf User0}'s number of Twitter friends (Log).
\item {\bf User0}'s number of Twitter followers (Log).
\item {\bf User0}'s number of Tweets (Log).
\item {\bf User0}'s number of favorite tweets (Log).
\item {\bf User0}'s total number of retweets (Log).
\item Average number of friends of {\bf User0}'s friends (Log).
\item Median number of friends of {\bf User0}'s friends (Log).
\item Standard deviation of the number of friends of {\bf User0}'s friends (Log).
\item Average number of followers of {\bf User0}'s friends (Log).
\item Median number of followers of {\bf User0}'s friends (Log).
\item Standard deviation of the number of followers of {\bf User0}'s friends (Log).
\item Average number of tweets of {\bf User0}'s friends (Log).
\item Median number of tweets of {\bf User0}'s friends (Log).
\item Standard deviation of the number of tweets of {\bf User0}'s friends (Log).
\item Average number of favorite tweets of {\bf User0}'s friends (Log).
\item Median number of favorite tweets of {\bf User0}'s friends (Log).
\item Standard deviation of the number of favorite tweets of {\bf User0}'s friends (Log).
\item Binary indicator of whether {\bf Friend}'s account authenticity had been verified by Twitter.
\item Fraction of {\bf User0}'s friends that had account authenticity verified by Twitter.
\item Binary indicator of whether {\bf Friend} and {\bf User0} had the same account language setting.
\end{itemize}

\end{APPENDIX}

\end{document}